\documentclass[11pt]{article}

\usepackage{amssymb}
\usepackage{graphicx}
\usepackage{color}
\usepackage{epstopdf}
\usepackage{amsmath,amsthm,amsfonts,epsfig,setspace,cite}

\newcommand\blfootnote[1]{%
  \begingroup
  \renewcommand\thefootnote{}\footnote{#1}%
  \addtocounter{footnote}{-1}%
  \endgroup
}

\title{Hybridization of Singular Plasmons via Transformation Optics}
\date{}

\author{Sanghyeon Yu\thanks{\footnotesize Department of Mathematics, ETH Z\"urich, R\"amistrasse 101, CH-8092 Z\"urich, Switzerland. }  \thanks{\footnotesize Correspondence to S.Y. (sanghyeon.yu@sam.math.ethz.ch)} \and Habib Ammari\footnotemark[1]}

\begin{document}
\maketitle

\begin{abstract}

Surface plasmon resonances of metallic nanostructures offer great opportunities to guide and manipulate light on the nanoscale. In the design of novel plasmonic devices, a central topic is to clarify the intricate relationship between the resonance spectrum and the geometry of the nanostructure. Despite the many advances, the design becomes quite challenging when the desired  spectrum is highly complex. Here we develop a new theoretical model for surface plasmons of interacting nanoparticles to reduce the complexity of the design process significantly. Our model is developed by combining plasmon hybridization theory with transformation optics, which yields an efficient way of simultaneously controlling both global and local features of the resonance spectrum. As an application, we propose a design of metasurface whose absorption spectrum can be controlled over a large class of complex patterns through only a few geometric parameters in an intuitive way. Our approach provides fundamental tools for the effective design of plasmonic metamaterials with on-demand functionality. \blfootnote{\textbf{Author Contribution}: S.Y. conceived the idea and designed research. S.Y. and H.A. performed research and wrote the paper.}

\end{abstract}

\medskip

\bigskip

Metallic nanostructures have been extensively studied and utilized for sub-wavelength control of light due to their unique ability to support surface plasmon resonances, which are the collective oscillations of conduction electrons on metal-dielectric interfaces \cite{SciRev, SLal, JAIN2010153, NatRev, fano, GB_NatPho_2010, AtwaterPolman,  Halas2011, new1, new2, Engheta1698, 
olson2014vivid,Kauranen,PendryTO2012,
PendryTO2015,
marinica2015active,
byers2015tunable,
brongersma2015plasmon, colour, yi2017vibrational, chapkin2018lifetime, zhou2018quantifying}. 
The excitation of surface plasmon resonances leads to the concentration of light into nanometric volumes and extreme enhancement of the electromagnetic fields. These phenomena have important applications including optical nanocircuits \cite{Engheta1698}, single molecule sensing, spectroscopy, light harvesting \cite{AtwaterPolman}, color nanotechnology \cite{colour}, and nonlinear optics \cite{Kauranen}. 

When designing plasmonic devices, one of the fundamental challenges is to find a geometry of the nanostructure which would yield the desired resonance spectrum. The Plasmon Hybridization (PH) model \cite{Prodan2003, Halas2011, Nordlander2004} has been successfully used to understand the spectral responses of various nanostructures in a simple and intuitive way. 
Since the  PH model guides us to design overall features of the nanostructures  with intuition, the specific characteristics can be optimized by tuning over small sets of structural parameters.
For designing  novel devices with custom-defined functionality, the desired spectra could be highly complex. In this case, continuing with the above direct design method, which is an intuition-based approach, encounters the challenge of increasing complexity.
Inverse design methods \cite{optimal1}, such as the adjoint method \cite{piggott2015inverse}, evolution algorithms \cite{johlin2018broadband}, or data-driven approaches based on machine learning 
\cite{peurifoy2018nanophotonic,
optimal2,
liu2018training} could be used instead but these require significant computational effort. It would be preferable to use a deeper theoretical understanding to reduce the complexity of the direct design problem.

To mitigate this challenge, we propose a new hybridization model for systems of strongly interacting particles by combining the PH model with Transformation Optics (TO) \cite{PendryTO2012,
PendryTO2015,TO1,TO2,TO3}, which is another theoretical method to understand surface plasmons. Our model greatly extends the applicability of the PH model so that it can describe a wide range of simple to complex spectrum patterns with only a few geometric parameters. In some sense, the TO approach captures the global behavior of the resonance spectrum, while the PH model captures the local spectral shift or splitting. Combining these two in an elegant way, our proposed model provides a deep physical insight into how to control the global and local features of the spectrum separately so that the range of achievable spectral patterns can be greatly extended. As a proof of concept, we propose a design of original metasurfaces whose absorption spectrum can be varied over a wide class, including  simple and complex patterns,  by changing only two geometric parameters in an intuitive manner. 
Our work shows the possibility of designing plasmonic metamaterials in an effective way even when the desired spectral response is highly complex.

\section*{Results and Discussions}

Before explaining our model, we briefly review the PH model and TO approach and discuss the related challenges.
In the PH model, the plasmons of the whole interacting particle system are viewed as simple combinations of the individual particle plasmons. 
The PH model describes the spectral shifts, induced by the coupling between the particles, in a way analogous to molecular orbital theory, providing a general and powerful design principle \cite{Prodan2003, Halas2011, Nordlander2004}. 
However, when the particles are extremely close-to-touching, the physical picture becomes complicated since a large number of uncoupled plasmons contribute to each hybridized plasmon.

Recently, the TO approach \cite{TO1,TO2,TO3} has been applied to two close-to-touching particles and other geometrically singular structures \cite{Pendry_twoparticles1, PendryTO2015, PendryTO2012, PendryTO2013, PendryTO2017}. We also refer to \cite{McPhedran1981, McP, McP_prsa_1988,poladian1989asymptotic, poladian1988general, Poladian1990, stout2008recursive,
lim2009blow, 
ammari2013spectral, bonnetier2013spectrum, 
schnitzer2015singular, sirev} for related works. TO shows that, as the two particles get closer, the discrete plasmon spectrum becomes more dense and eventually converges to a continuous spectrum at the touching limit. 
In the TO approach, conformal mappings are used to transform the singular structure  to one with the same spectrum but having nicer symmetry, thereby providing a unique physical insight into the origin of the broadband light harvesting as well as analytic  solutions (SI Appendix, sections 1-3). 
 Moreover, for two close-to-touching particles, the TO approach is more accurate and efficient than the standard hybridization method (see SI Appendix, section 4 and Fig. S2 for their comparison).
 Nevertheless,  TO alone cannot be applied to systems featuring three (or more) particles.

We should mention that, in our previous work \cite{sirev}, we considered a similar problem and developed an efficient numerical method for a system of close-to-touching plasmonic particles.  But this numerical method, like FEM or FDTD, does not provide deep physical insights into the hybridization of surface plasmons. We emphasize that our present work's focus is on advances in physical understanding of the plasmonic interaction.

\begin{figure}
\centering
\includegraphics[width=0.70\linewidth]{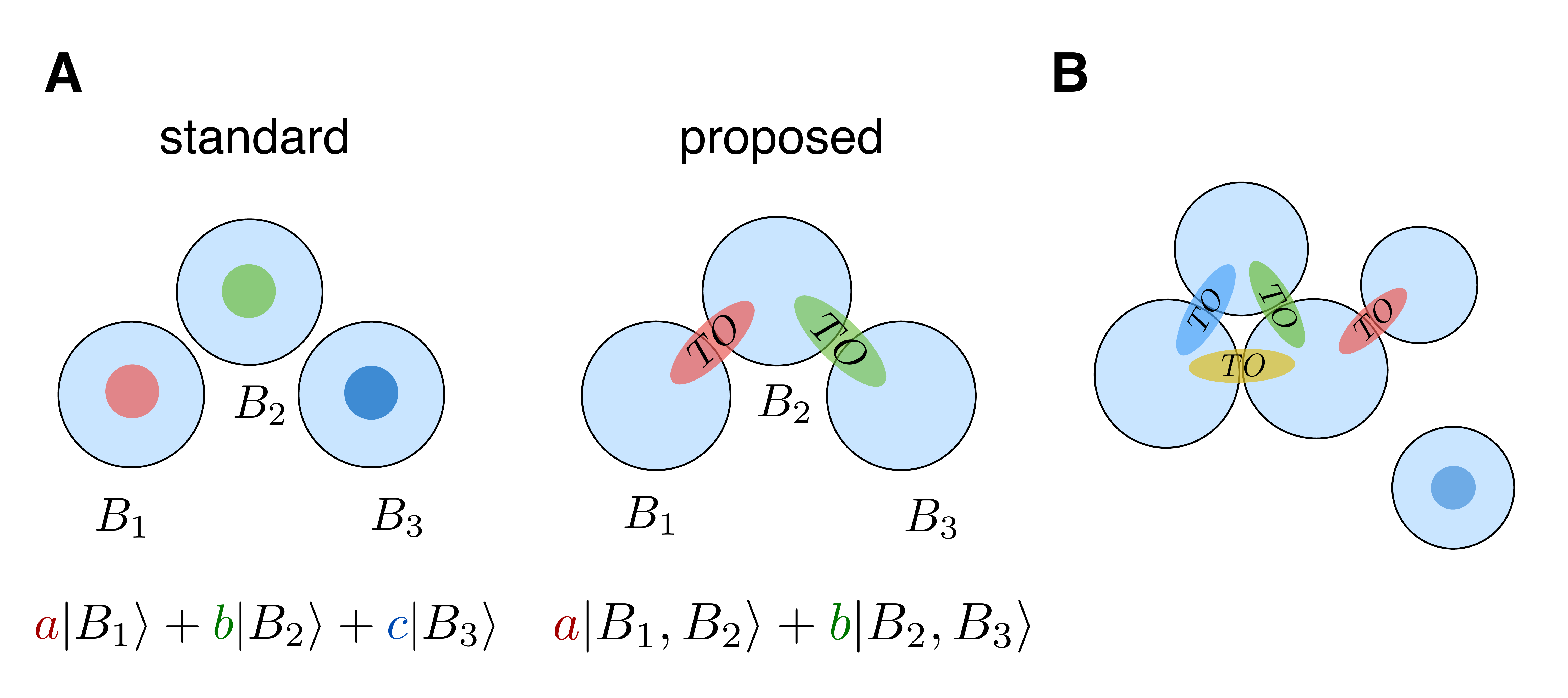}
\caption{ Schematic description of our proposed model. {(A)} comparison with the standard hybridization model. {(B)} a more general case.}
\label{fig:trimer_geom}
\end{figure}

As mentioned previously, our proposed model combines the advantages of both the PH and TO approaches to deal with an arbitrary number of close-to-touching particles. 
We shall consider the system of close-to-touching particles as a prototypical example. Our approach can more generally be applied to other singular systems consisting of crescents with corners, eccentric shells with small gaps or their mixtures.

We should mention that the non-local effect, which has a quantum origin, is an important issue when the gap distance between the particles is extremely small (below $0.25$ nm) \cite{prod_nonlocal, Ciraci2012_nonlocal, Zhu2016_nonlocal, Luo2013_nonlocal, Luo2014_nonlocal, ory_nonlocal}. Our focus is {\sl not} on modelling the non-local effect but on understanding the strong interaction between the particles. We shall assume a local model for the metal permittivity. The nonlocal effect can be accounted for by using the approach of \cite{Luo2013_nonlocal,Luo2014_nonlocal,PendryTO2012}.

We now explain our proposed model which we call the {\it Singular Plasmon Hybridization (SPH) Model}. In the standard hybridization model, a plasmon of the system is a combination of plasmons of individual particles. On the contrary, in our approach, the basic building blocks are the gap-plasmons of a pair of particles whose singular behavior is captured using the TO approach. This simple conceptual change is the key to solving the aforementioned challenges. In Fig. 1A, we show a schematic comparison for a trimer, as it is the simplest example for our model (we emphasize that our model can be applied to a general configuration of particles, as shown in Fig. 1B). The trimer plasmon is now treated as a combination of two gap-plasmons. In our picture, the new plasmons are formed by the hybridization of these gap-plasmons. The gap-plasmons are strongly confined in their respective gaps and all the gaps are well-separated meaning that the gap-plasmons do not overlap significantly. Hence, the spectral shifts due to their hybridization should be moderate and we can expect to find a simple picture even in the close-to-touching case.

To gain a better understanding, we develop a coupled mode theory for the hybridization of singular gap-plasmons. For simplicity, we consider only two-dimensional structures (however our theory can be extended to the three-dimensional case). We assume the Drude model for the metal permittivity 
$\epsilon = 1- {\omega_p^2}/{\omega^2}$,
where $\omega_p$  is the bulk plasma frequency and the background permittivity is $\epsilon_0=1$. We also adopt the quasi-static approximation by assuming the system to be small compared to the wavelength of the incident light. We remark that the radiation reaction can be incorporated to go beyond the quasi-static limit, as described in \cite{PendryTO2012, aubry2010conformal}.

\begin{figure*}
\centering
\includegraphics[width=.55\linewidth]{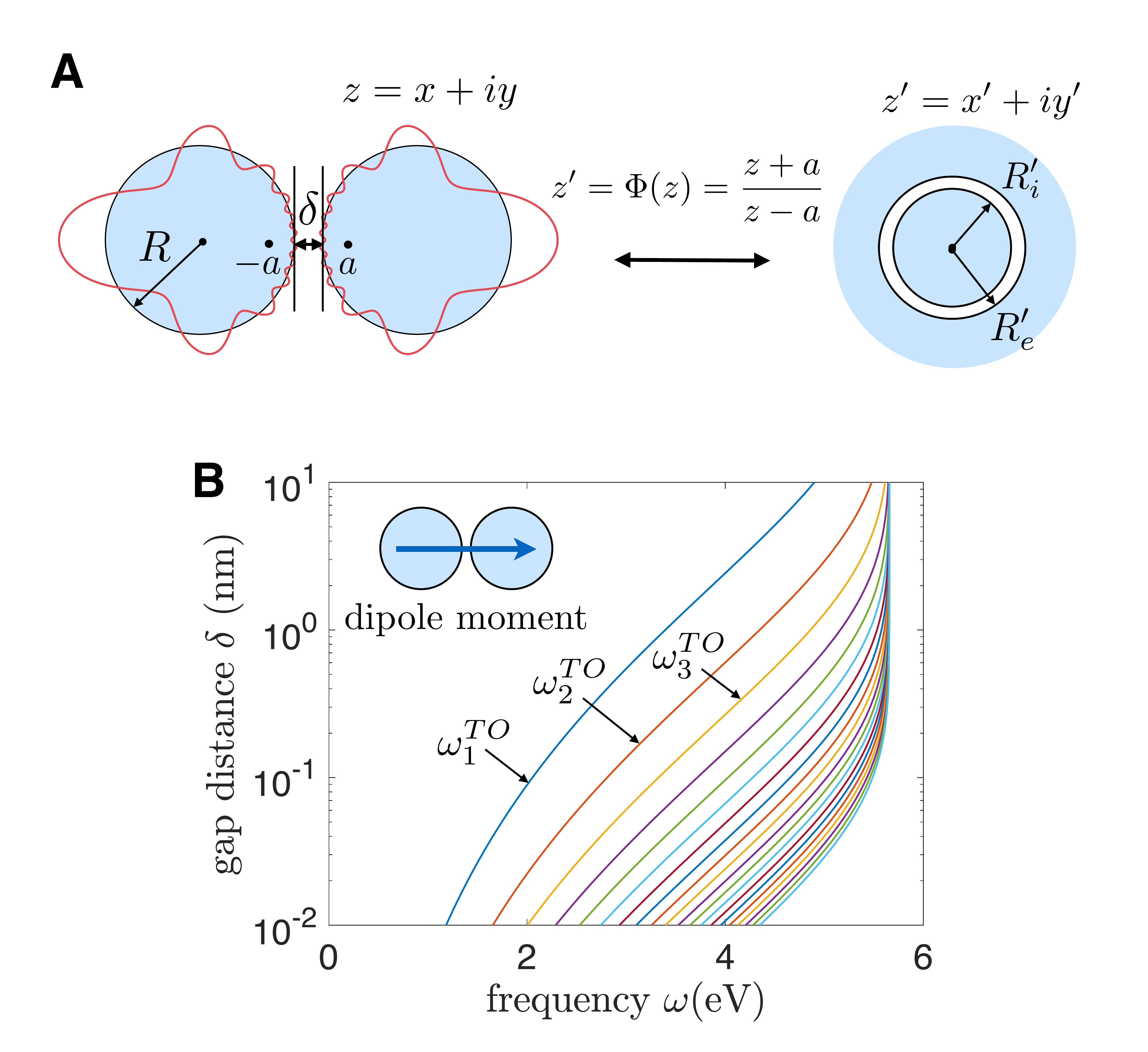}
\caption{{Dimer Plasmons in TO approach.} {(A)} Transformation of a dimer into a concentric shell. Red solid lines represent the oscillations of a dimer 
gap-plasmon. {(B)} The singular red-shift of the spectrum for a dimer. We set $R=20$  nm and $\omega_p = 8$ eV. 
}
\label{fig:dimer_TO}
\end{figure*}

We begin with the TO description \cite{Pendry_twoparticles1,PendryTO2012,PendryTO2013,PendryTO2015,PendryTO2017} of gap-plasmons which are the basic building blocks of our proposed model. Consider a dimer of cylinders of radius $R$ separated by a distance $\delta$. By an inversion conformal mapping, the dimer is transformed to a concentric shell which is an analytically solvable case (Fig. \ref{fig:dimer_TO}A).  
Then TO  reveals that, when the two cylinders get closer, the wavelength of their plasmon near the gap becomes smaller and energy accumulation occurs in the gap region. This gives rise to an extreme field enhancement. TO also can describe the singular spectral shift of gap-plasmons. Let us consider the gap-plasmons whose dipole moment is aligned parallel to the dimer axis since these plasmons contribute to the optical response significantly. Their resonant frequencies $\omega_n^{TO}$ are given by
\begin{equation}
\omega_n^{TO} = {\omega_p}/\sqrt{e^{-ns}\sinh (ns)}, \quad n=1,2,3,\cdots,
\end{equation}
with the parameter $s$ satisfying $\sinh^2 s = (\delta/R)(1+\delta/4R)$. We denote their associated gap-plasmons by $|\omega_n^{TO}\rangle$. When the gap distance $\delta$ gets smaller, as shown in Fig. 2b, the frequencies $\omega_n^{TO}$ are red-shifted singularly and the spectrum becomes denser. Thus, the TO description captures the singular behavior of gap-plasmons (see SI Appendix, section 2 for more details).

\begin{figure*}
\centering
%\includegraphics[width=.15\linewidth]{trimer_geometry}
%\includegraphics[width=.15\linewidth]{trimer_hybridization}
%\vskip.05cm
%\includegraphics[width=.25\linewidth]{absor_trimer_smallgap}
%\includegraphics[width=.25\linewidth]{absor_trimer_extremegap}
\includegraphics[width=.90\linewidth]{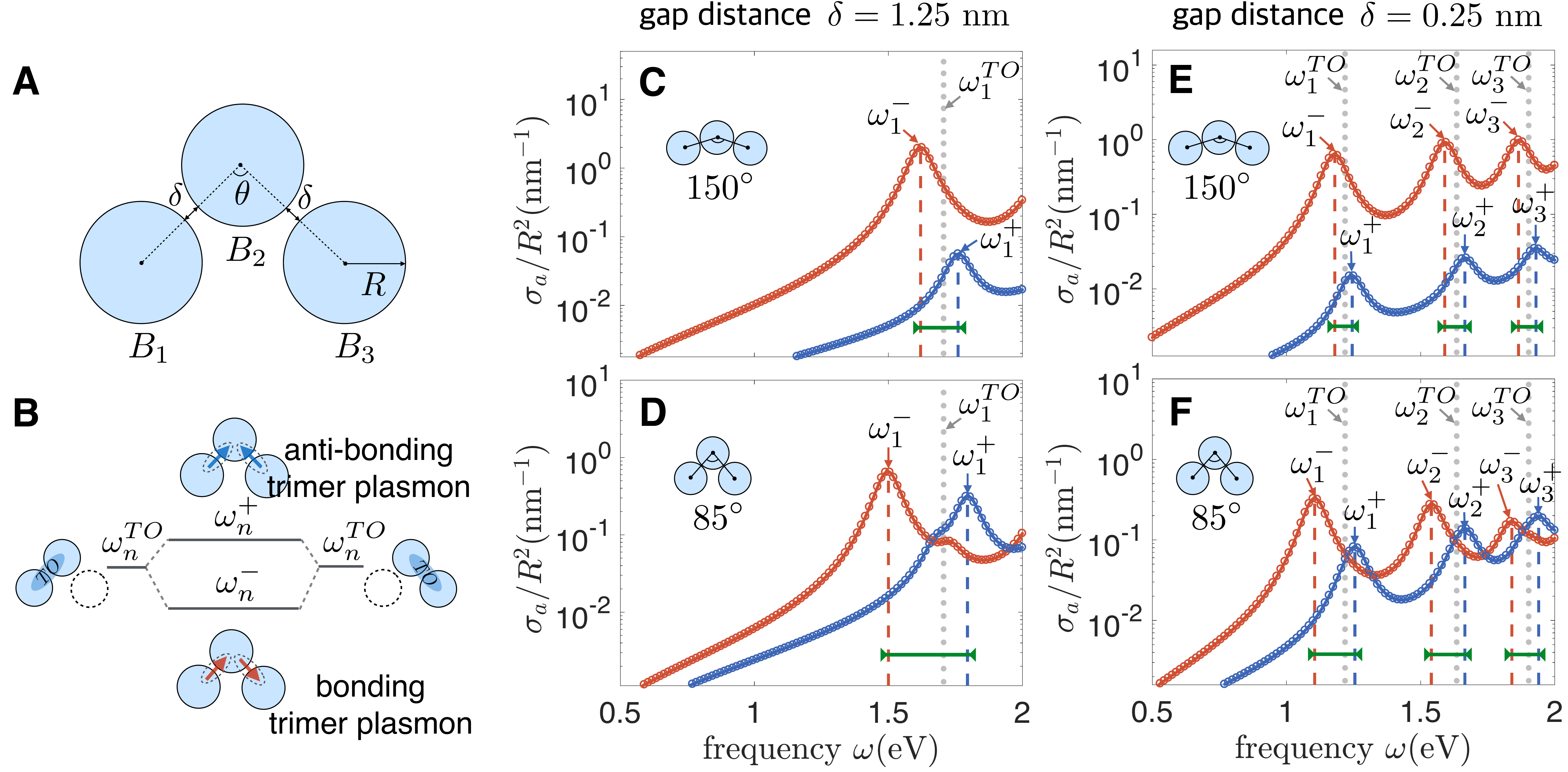}
\caption{ 
Trimer plasmons in SH model. 
(A) Geometry of the trimer. The pairs $(B_1,B_2)$ and $(B_2,B_3)$ are close-to-touching while $B_1$ and $B_3$ are well-separated. {(B)} Hybridization of trimer singular plasmons.
(C,D) Absorption cross sections when the gap distance is $\delta=1.25$ nm. 
The open circles (resp. solid lines) represent the numerical simulation results (resp. theoretical results from the SPH model), respectively.
The red and blue colors represent the x-polarized and y-polarized incident light cases, respectively.
(E,F) The same as (C,D) but with the gap distance $\delta=0.25$ nm. We set $R=20$ nm, $\omega_p = 3.85 $ eV and $\gamma = 0.1 $ eV.
}
\label{fig:trimer_abs}
\end{figure*}

We now turn to our SPH model, taking a trimer as an example (Fig. \ref{fig:trimer_abs}A). The trimer plasmon is specified as a superposition of the gap-plasmon of the pair $(B_1,B_2)$ and that of the pair $(B_2, B_3)$.
We let $(a_n, b_n)$ represent the following linear combination of the gap-plasmons: $a_n |\omega_{n}^{TO}(B_1,B_2) \rangle + b_n |\omega_{n}^{TO}(B_2,B_3) \rangle  $.
Their hybridization is characterized by the following coupled mode equations:
\begin{equation}
\begin{bmatrix}
(\omega_n^{TO})^2 & \Delta_{n}
\\
\Delta_{n} & (\omega_n^{TO})^2 
\end{bmatrix}
\begin{bmatrix}
a_n
\\
b_n
\end{bmatrix}
=
\omega^2
\begin{bmatrix}
a_n
\\
b_n
\end{bmatrix}.
\end{equation}
Here, $\Delta_n$ represents the coupling between the two gap-plasmons.
As the bonding angle $\theta$ between  the two gap-plasmons decreases, the coupling strength $\Delta_n$ increases, which is to be expected since the two gaps get closer. 
This coupled mode system is derived using the spectral theory of the Neumann--Poincar\'e operator \cite{Ammari_2016_NP,arma2017,jde2016,AK_2016_NP,Mayer2005_NP, M11} and TO (see SI Appendix, section 5 for details).  The coupling term $\Delta_n$ is calculated without any assumption or fitting procedure.
We emphasize that the above equation is a simplified version of our theory. Although we require additional TO gap-plasmons for improved accuracy, we shall see that this simplified version can already capture the physics.  
Solving the equation, we obtain the hybridized plasmons for the trimer
\begin{equation}
|\omega_n^{\pm}  \rangle  \approx  \frac{1}{\sqrt{2}}\Big(|\omega_{n}^{TO}(B_1,B_2) \rangle \mp |\omega_{n}^{TO}(B_2,B_3) \rangle\Big), \quad n=1,2,\cdots,
\end{equation}
and their resonant frequencies 
\begin{equation}
\omega_n^{\pm} \approx \omega_n^{TO} \pm \Delta_n, \quad n=1,2,\cdots.
\end{equation}
So our theory predicts that the spectrum consists of a family of pairs $(\omega_n^-,\omega_n^+)$ of resonant frequencies which are split from the dimer resonant frequencies $\omega_n^{TO}$.
The dimer part $\omega_n^{TO}$ is singularly shifted as the gap distance $\delta$ gets smaller, while the splitting part $\Delta_n$ remains moderate.

We call $|\omega_n^-\rangle$ and $|\omega_n^+\rangle$  the {\it bonding trimer plasmon} and {\it anti-bonding trimer plasmon}, respectively (Fig. \ref{fig:trimer_abs}B). 
The bonding plasmon has a net dipole moment pointing in the x-direction so that it can be excited by the x-polarized light. Similarly, the anti-bonding plasmon can be excited by the y-polarized light. 
These plasmons are very different from the bonding plasmon and anti-bonding plasmon of a dimer in the standard hybridization model. They are trimer plasmons and are capable of capturing the close-to-touching interaction via TO. 
We emphasize that, contrary to the standard hybridization approach, these `simple' combinations of gap-plasmons remain effective for describing the hybridized plasmons even when the particles are close-to-touching. In other words, the required number of uncoupled gap-plasmons does not increase as the gap distance $\delta$ gets smaller and hence our model gives a simple picture in the close-to-touching case.
We also mention that our physical picture for the trimer is qualitatively different from the standard hybridization one given in \cite{trimer1,trimer2}.

 We now discuss how our SPH model gives new physical insights into the relationship between geometry and plasmons. The power of our model comes from its ability to decompose the plasmon spectrum into a singular part, which depends on the local geometry, and a regular part, which depends on the global geometry. 
The resonant frequency $\omega_n^{\pm}$ for the trimer consists of two parts: the singularly shifted part $\omega_n^{TO}$ and the regular splitting part $\Delta_n$. 
The singular part $\omega_n^{TO}$ is determined by the small gap distance $\delta$, which is a {`local'} feature of the geometry. On the other hand, the regular part $\Delta_n$ is determined by the bonding angle $\theta$, which is a `global' feature of the geometry.
In other words, the small gap distance $\delta$ affects the `overall' behavior of the spectrum while the bonding angle $\theta$ controls the `detailed' splitting of the spectrum. This shows an interesting relation between the spectrum and the geometry:  {\sl local} (and {\sl global}) features of the geometry can determine the {\sl global}  (and {\sl local}) behavior of the spectrum, respectively. 
This relation provides us with a design principle: one should manipulate the local geometric singularity (such as inter-particle gap distances) to control the overall behavior of the spectrum, while manipulating the global geometry to achieve the desired detailed splitting of the spectrum.
Our SPH model provides a systematic way of achieving such a design using gap-plasmons as basic building blocks. 
Our approach is valid for general systems consisting of an arbitrary number of interacting particles, with arbitrary positions and different radii.

We validate our model with numerical examples for the trimer. We set the radius of the particles to be $R=20$~nm. We consider the two cases: 
{when the inter-particle gap distances are (i) $\delta = 1.25$~nm and (ii) $\delta = 0.25$~nm}. Notice that the ratio $\delta/R$ is very small so that the particles are close to touching. We assume the Drude model $\epsilon=1-\omega_p^2/(\omega(\omega+i\gamma))$ with $\omega_p=3.85$ eV and $\gamma = 0.1$ eV.
In Figures \ref{fig:trimer_abs}C and \ref{fig:trimer_abs}D, we plot the absorption cross section (red and blue  circles) for the trimer with  the gap distance $\delta=1.25$ nm when the bonding angle  is $\theta=150^\circ$ (weak coupling)  and $\theta=85^\circ$ (strong coupling), respectively. In the latter case, the coupling strength between the gap-plasmons  is stronger since the gaps are  closer to each other. 
The solid lines (and open circles) represent the theoretical results (and numerical simulation results) performed by our SPH model  (and the fully retarded version of multipole expansion method), respectively. 
The red and blue colors correspond to the x-polarized and y-polarized incident light cases, respectively.
Similarly, in Figs. \ref{fig:trimer_abs}E and \ref{fig:trimer_abs}F, we plot the absorption cross section in the case of the smaller gap distance $\delta=0.25$ nm. 
We emphasize that we have used a complete version of our SPH model for theoretical results (see SI Appendix, section 5 for details).
 We also plot the values of the resonant frequencies $\omega_n^-$ and $\omega_n^+$ (red and blue, dashed, vertical lines) computed by the SPH model.
 Their corresponding plasmons are dominated by bonding and anti-bonding combinations of gap-plasmons, respectively. 
 As expected, the resonance peaks of the absorption are located near the bonding (and anti-bonding) plasmon frequencies $\omega_n^-$ (and $\omega_n^+$) for the x-polarized (and y-polarized) incident field, respectively. See SI Appendix, section 6 and Figs. S3-6 for the contour plots of potential distributions at these resonant frequencies.
  The gray dots represent the dimer frequency $\omega_n^{TO}$ computed using the TO approach.
As the gap-distance $\delta$ gets smaller, the overall spectrum is significantly red-shifted in conjunction  with the singular shift of $\omega_n^{TO}$.
 The green arrows indicate how much the trimer frequencies $\omega_n^\pm$ have split from the dimer frequency $\omega_n^{TO}$.
It is clearly shown that the splitting $\omega_n^{\pm}-\omega_n^{TO}$ is moderate regardless of the inter-particle gap-distance. In the strong coupling case (smaller bonding angle) the splitting is more pronounced. Further, as the bonding angle $\theta$ decreases the absorption of the y-polarized incident field becomes stronger since the net dipole moment of the anti-bonding mode increases. 
Hence, the numerical results are consistent with the prediction of our proposed SH model.

\begin{figure*}
\centering
\includegraphics[width=0.95\linewidth]{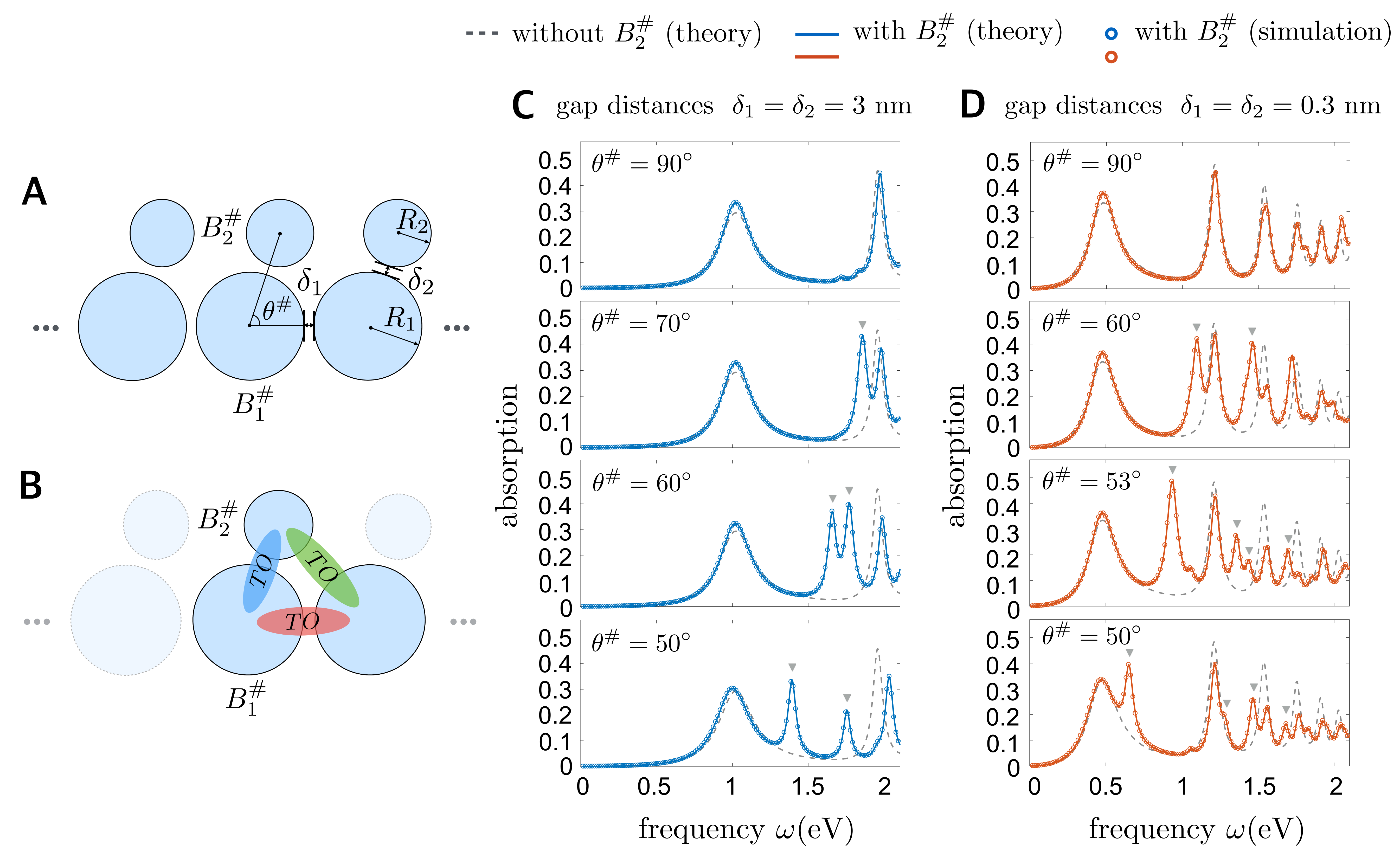}
\caption{Metasurface geometry (A), singular gap-plasmons of the metasurface (B) and the absorption spectrum patterns (C,D) for various gap distances and bonding angles . We set $R_1=30$ nm, $R_2=15$ nm, $\omega_p = 3.5 $ eV and $\gamma = 0.05 $ eV. }
\label{fig:metasurface}
\end{figure*}

We now move on to consider the design of metasurfaces, which is a key application of this work.
Recently, Pendry et al. \cite{PendryTO2017} proposed a broadband absorbing metasurface consisting of a grating with points of vanishingly small thickness (which are geometric singularities). Remarkably, they interpreted its broadband spectral response as a realization of compacted dimensions. 
They used a geometric singularity to control the global behavior of the spectrum: the discrete spectrum converges to the continuous one as the small thickness goes to zero.  We refer to \cite{PendryPRX2015, PendryPRB2018, PendryPRB2019} for related works on the metasurfaces.
In this work, we propose a metasurface which can generate a large class of simple to complex spectral patterns through only two geometric parameters. This is achieved by controlling both global and local spectral shifts. 
Our metasurface geometry, shown in Figure \ref{fig:metasurface}A, is a one-dimensional periodic array consisting of two particles with different radii. This array is a combination of two sub-arrays, 
 $B_1^\#$ which consists of the largest particles, and  $B_2^\#$ which consists of the smallest ones.

To explain the motivation behind our metasurface design, we begin by considering the array of larger particles $B_1^\#$. 
As the particles comprising $B_1^\#$ become closer, the spectrum becomes more dense. 
To generate various spectral patterns, we introduce the array  $B_2^\#$ and position it close to $B_1^\#$. This leads to the formation of three kinds of singular gap-plasmon arrays in each gap, as in Figure \ref{fig:metasurface}B, the hybridization of which results in new resonance peaks. As in the case of a single trimer, it is natural to expect that the hybridization becomes stronger  as the bonding angle $\theta^\#$ decreases. 
We verify this prediction with numerics.
We set the radii to be $R_1=30$~nm, $R_2=15$~nm. We assume that the two gap distances are the same, $\delta_1=\delta_2=\delta$, and consider two cases: (i) $\delta=3$~nm and (ii) $\delta= 0.3$~nm. 
For each gap distance, we also consider four cases of different bonding angles from $\theta^{\#}= 90^\circ$ (weak coupling)  to $\theta^{\#}=50^\circ$ (strong coupling).
We plot the absorption in Figures \ref{fig:metasurface}C and \ref{fig:metasurface}D, respectively, assuming a normal incidence to the metasurface. Assuming the Drude model $\epsilon=1-\omega_p^2/(\omega(\omega+i\gamma))$ with $\omega_p=3.5$ eV and $\gamma = 0.05$ eV,
we compute the absorption spectra using our SPH model. See SI Appendix, section 8 for  the SPH theory of the metasurface.
Our theoretical results, which are based on the quasi-static approximation, are in excellent agreement with the fully retarded simulation results (performed by using the multipole expansion method). This is because the period of the structure is small compared to the wavelength.
We also compare the results with the absorption of a metasurface consisting of only the largest particles $B_1^\#$ (gray dotted lines). 
In the weak coupling regime, the introduction of the array  $B_2^\#$ results in a somewhat minor alteration of the spectrum.
In the strong coupling regime, new resonance peaks appear due to the strong hybridization of singular plasmons (the new peaks are marked with gray triangles). Clearly, this shows that a variety of patterns of the plasmon spectrum, ranging from simple to highly complex ones,  are generated by adjusting only two geometric parameters: the gap distance $\delta$ and the bonding angle $\theta^\#$. This is possible because $\delta$ controls the global spectral shift while  $\theta^\#$ controls the local spectral splitting. The diversity of the generated spectrum comes from a combination of these two geometric parameters.
This could have applications in plasmonic color engineering \cite{olson2014vivid, colour}.  
We should also mention that these various spectral patterns and their resonance peaks can be explained using our SPH model, although we shall not present the detailed analysis for the hybridized modes here. More general spectral patterns could be realized by considering more complex particle configurations.

Let us discuss the advantages of the SPH model gives us in the design of the metasurface. First, the proposed metasurface has `only a few' geometric parameters but can generate a large class of spectral responses. 
Second, the SPH model yields (semi) `analytic' solutions for the hybridized plasmons of the metasurface, as described in the SI Appendix, section 8. These two features allow us to easily optimize the metasurface's parameters for the targeted application.

We finally mention that the extension of the SPH model to the 3D case is non-trivial. A spherical dimer supports three branches of gap plasmons \cite{PendryTO2013}: the odd modes, normal even modes and anomalous even modes. The anomalous ones are not confined entirely within the gap but spread over the whole particle surface. Hence their hybridization cannot be described by the present form of the SPH model and should be considered carefully.

\section*{Conclusions}

In conclusion, we have proposed the {\it Singular Plasmon Hybridization Model} for plasmons of strongly interacting particles which gives a simple and intuitive physical picture when the particles are close-to-touching. The proposed model demonstrates an elegant interplay between the plasmon hybridization model and transformation optics, clarifying a deep geometric dependence of the plasmon spectrum.
It enables us to design a plasmonic device whose spectral characteristics can be controlled over a large class of patterns through only a few geometric parameters.
We believe that our model can have a significant impact on the design of a variety of complex plasmonic devices.

% Bibliography
\newpage

\section*{Supplementary Materials}

Here we describe our coupled mode theory for the hybridization of singular gap-plasmons.
We consider the 2D case for simplicity. We also assume the quasi-static approximation.

\section{Integral equation approach for surface plasmons}
Suppose we have a system of nanoparticles $\Omega$ with permittivity $\epsilon$. We assume that the background permittivity is $\epsilon_0=1$ and denote by $\mathbf{E}^{in}$  the incident electric field.
Let $G$ be the electric potential generated by a unit point charge in 2D (or a line charge in 3D), {\it i.e.}, 
\begin{equation}
    G(\mathbf{r}) = -\frac{1}{2\pi}\ln |\mathbf{r}|, \quad \mathbf{r}\in \mathbb{R}^2\setminus \{(0,0)\}.
\end{equation}
Then the induced charge density $\sigma$ on the surface $\partial \Omega$ of the system is determined by the following integral equation \cite{Ammari_2016_NP, AK_2016_NP, Mayer2005_NP}:
\begin{equation}\label{main_integral_eqn}
(\mathcal{K}^*_\Omega - \lambda I)[\sigma] = \mathbf{E}^{in}\cdot \mathbf{n}|_{\partial\Omega}, \quad \lambda = \frac{\epsilon+1}{2(\epsilon-1)},
\end{equation}
where $\mathcal{K}^*_\Omega$ is the Neumann--Poincar\'e (NP) operator  given by 
\begin{equation}
\mathcal{K}^*_\Omega[\sigma] (\mathbf{r}) = -\int_{\partial\Omega} \frac{\partial G(\mathbf{r}-\mathbf{r}')}{\partial \mathbf{n}(\mathbf{r})} \sigma(\mathbf{r}')dS(\mathbf{r}'), \quad \mathbf{r}\in \partial\Omega,
\end{equation}
and $\mathbf{n}$ is the outward unit  normal vector to the surface. If the permittivity $\epsilon$ is negative, then the above problem may admit a solution even when the incident field $\mathbf{E}^{in}$ is absent. In fact, this solution corresponds to the (localized) surface plasmon of the given system. More precisely, the mathematical analysis of the  surface plasmons is equivalent to the following eigenvalue problem for the NP operator \cite{Ammari_2016_NP, AK_2016_NP, Mayer2005_NP}:
\begin{equation}
 \mathcal{K}^*_\Omega[\sigma] = \lambda \sigma, \quad \lambda = \frac{\epsilon+1}{2(\epsilon-1)}.
\end{equation}
If we model the metal permittivity  by the lossless Drude's model in which $\epsilon = 1-{\omega_p^2}/{\omega^2},$ then the above eigenvalue problem can be rewritten as
\begin{equation}
\mathcal{A}_\Omega [\sigma] : =  \omega_p^2 \Big(\frac{1}{2}I-\mathcal{K}_\Omega^*\Big) [\sigma]= \omega^2 \sigma.
\end{equation}
Let $\omega_n^2$ and $\sigma_n$ be the eigenvalues and eigenfunctions of the operator $\mathcal{A}_\Omega$. 
Then $\omega_n$ (and $\sigma_n$) represents the resonance frequency (and the charge density) of plasmons, respectively. 
Let us denote the plasmon charge density $\sigma_n$  by $|\omega_n\rangle$ to indicate that its resonance frequency is $\omega_n$.
Note that the eigenvalues $\lambda_n$ of the operator $\mathcal{K}^*_\Omega$ are related to $\omega_n^2$ by $\lambda_n=1/2 - \omega_n^2/\omega_p^2$.

Let us define an inner product $\langle \omega_n | \omega_{n'} \rangle$ of two plasmons $|\omega_n\rangle$ and $|\omega_{n'}\rangle$ by 
\begin{equation}
\langle \omega_n | \omega_{n'} \rangle = \int_{\partial \Omega}\sigma_n(\mathbf{r})\int_{\partial \Omega} \frac{(-1)}{2\pi}\log|\mathbf{r}-\mathbf{r}'| \sigma_{n'}(\mathbf{r}') dS(\mathbf{r}')dS(\mathbf{r}).
\end{equation}
It can be shown that the eigenfunctions of the operator $\mathcal{K}^*_\Omega$ (hence $\mathcal{A}_\Omega$) form a complete orthogonal basis with respect to the above inner product.

Now we consider the Drude model with a small damping parameter $\gamma$, namely, $\epsilon = 1-\omega_p^2/(\omega(\omega+i\gamma))$. Then, by applying the spectral decomposition $\mathcal{K}_\Omega^* = \sum_{n=1}^\infty \lambda_n | \omega_n\rangle \langle \omega_n|$ to Eq. \ref{main_integral_eqn}, we obtain
\begin{equation}\label{spectral_decomp}
    \sigma = \sum_{n=1}^\infty \frac{\langle  \mathbf{E}^{in}\cdot \mathbf{n}|_{\partial \Omega}|\omega_n\rangle}{\lambda_n - \lambda} |\omega_n\rangle = \sum_{n=1}^\infty \frac{\omega_p^2\langle  \mathbf{E}^{in}\cdot \mathbf{n}|_{\partial \Omega}|\omega_n\rangle}{\omega^2 - \omega_n^2 + i \gamma \omega} |\omega_n\rangle.
\end{equation}
Then we can get the polarizability $\mathbf{p}$, which is the dipole moment of the charge density $\sigma$. The absorption cross section $\sigma_a$ (in the quasi-static approximation) can be computed by
\begin{equation}\label{absor}
\sigma_a = \omega\frac{ \mbox{Im}\{\mathbf{E}^{in}\cdot \mathbf{p}\} }{|\mathbf{E}^{in}|^2}.
\end{equation}

\section{TO description of the dimer plasmons}
Consider the dimer $D = B_+ \cup B_-$ where $B_\pm$ is a circular cylinder of radius $R$ centered at $\pm (R+\delta/2,0)$. Note that the two particles $B_+$ and $B_-$ are separated by a distance $\delta$. 
Using the TO approach \cite{Pendry_twoparticles1,PendryTO2012,PendryTO2013,PendryTO2015,PendryTO2017}, we can derive the dimer plasmons ({\it{i.e.}} the eigenvalues and eigenfunctions of the operator $\mathcal{A}_D$) explicitly. 
The conformal transformation $\Phi$ given by 
\begin{equation}
x'+i y' = \Phi(x+i y) = \frac{x+iy+a}{x+iy-a}, \qquad a = ({\delta (R+ \delta/4)})^{1/2},
\end{equation}
maps the dimer to a concentric annulus whose inner radius is $r_i = e^{-s}$ and outer radius is $r_e = e^{s}$, where $\sinh s= a/R$. 
Let $(r',\theta')$ be the polar coordinates of the transformed frame, namely,
$
z' = x'+iy' = r'e^{i\theta'}
$.
As mentioned in the main manuscript, we consider only the dimer plasmons whose dipole moment is aligned in the $x$-direction. The resonance frequencies of these plasmons are
\begin{equation}
\omega_n^{TO} = {\omega_p}/\sqrt{e^{-ns}\sinh (ns)}, \quad n=1,2,3,\cdots,
\end{equation}
and their associated plasmon charge densities $|\omega_n^{TO}\rangle$ are given as follows: for $n=1,2,3,\cdots$,
\begin{equation}
|\omega_n^{TO}(B_-,B_+) \rangle (\mathbf{r})  = \pm \frac{1}{\sqrt{N_n}} \frac{\cosh s-\cos\theta'}{\alpha} \cos {n\theta' }, \quad \mathbf{r}\in {\partial B_\pm },
\end{equation}
where the normalization constant  $N_n$ is chosen such that $\langle \omega_n^{TO}| \omega_n^{TO}\rangle = 1$. 
We can verify that $|\omega_n^{TO}\rangle$ are the eigenfunctions of $\mathcal{A}_D$ with the eigenvalues $(\omega_n^{TO})^2$, namely,  $\mathcal{A}_D|\omega_n^{TO}\rangle = (\omega_n^{TO})^2 |\omega_n^{TO}\rangle$.

Next we consider another branch of the dimer plasmons whose dipole moment is aligned in the $y$-direction. Their resonance frequencies are 
\begin{equation}
    \widetilde{\omega}_n^{TO} = \omega_p / \sqrt{e^{-ns} \cosh (ns)}, \quad n=1,2,3,\cdots,
\end{equation}
and the associated plasmon charge densities $|\widetilde{\omega}_n^{TO} \rangle$ are given as follows: for $n=1,2,3,\cdots,$ 
\begin{equation}
|\widetilde{\omega}_n^{TO}(B_-,B_+) \rangle (\mathbf{r})  =  \frac{1}{\sqrt{\widetilde{N}_n}} \frac{\cosh s-\cos\theta'}{\alpha} \sin {n\theta' }, \quad \mathbf{r}\in {\partial B_\pm },
\end{equation}
where $\widetilde{N}_n$ is the normalization constant.  As before, we have $\mathcal{A}_D|\widetilde{\omega}_n^{TO}\rangle = (\widetilde{\omega}_n^{TO})^2 |\widetilde{\omega}_n^{TO}\rangle$.

Note that $|{\omega}_n^{TO}(B_-,B_+) \rangle$ (resp. $|\widetilde{\omega}_n^{TO}(B_-,B_+) \rangle$) is anti-symmetric (resp. symmetric) with respect to the $y$-axis. So we shall refer to $|\widetilde{\omega}_n^{TO}(B_-,B_+) \rangle$ (resp. $|\widetilde{\omega}_n^{TO}(B_-,B_+) \rangle$) as the anti-symmetric (resp. the symmetric) dimer plasmons. 
It is worth mentioning that the resonance frequencies $\omega_n^{TO}$ (resp. $\widetilde{\omega}_n^{TO}$) lie below (resp. above) the surface plasmon frequency $\sqrt{2}\omega_p$.

\section{Potential distributions of the dimer plasmons}
\begin{figure*}
\centering
\includegraphics[width=8.5cm]{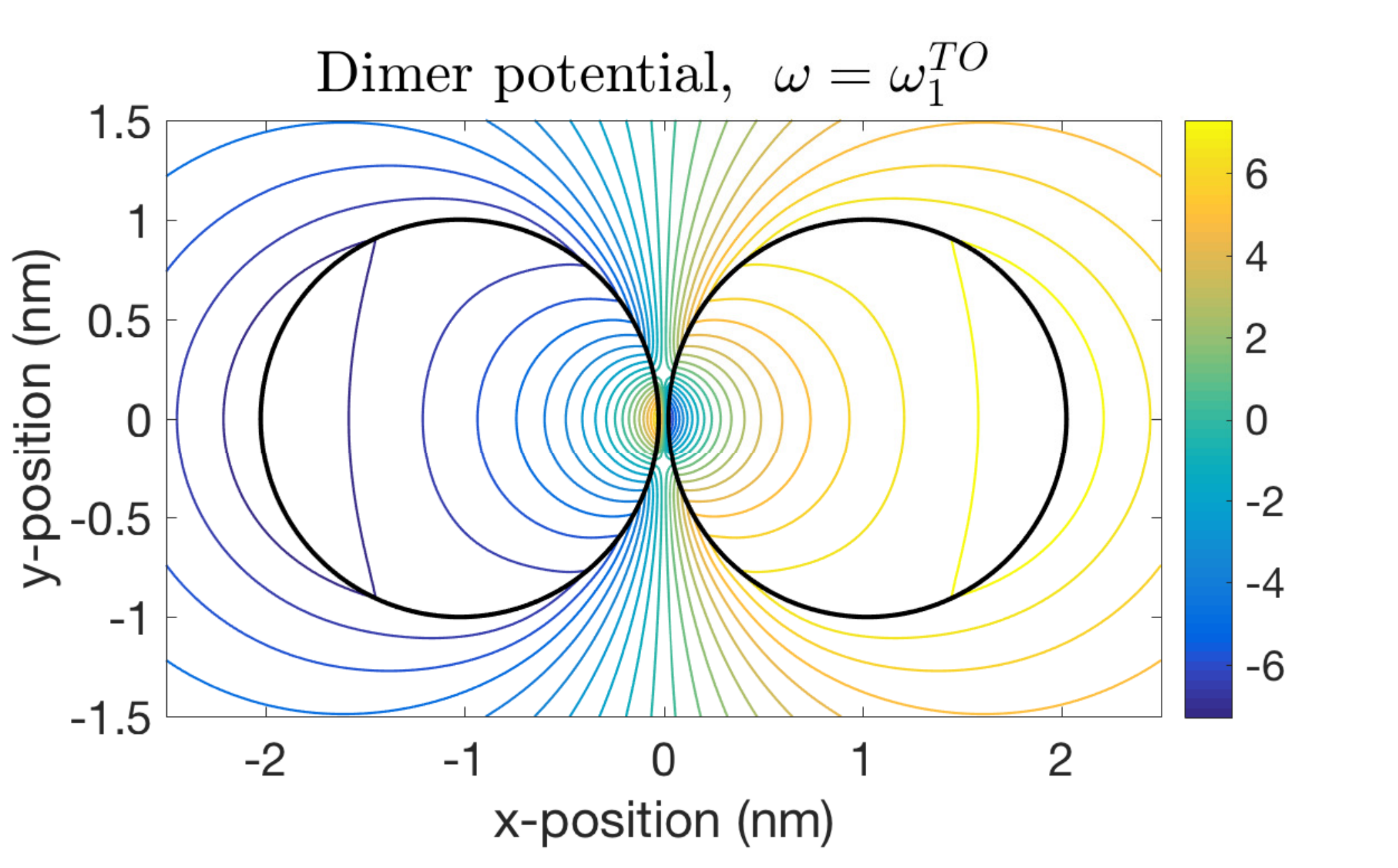}
\\[0.5em]
\includegraphics[width=8.5cm]{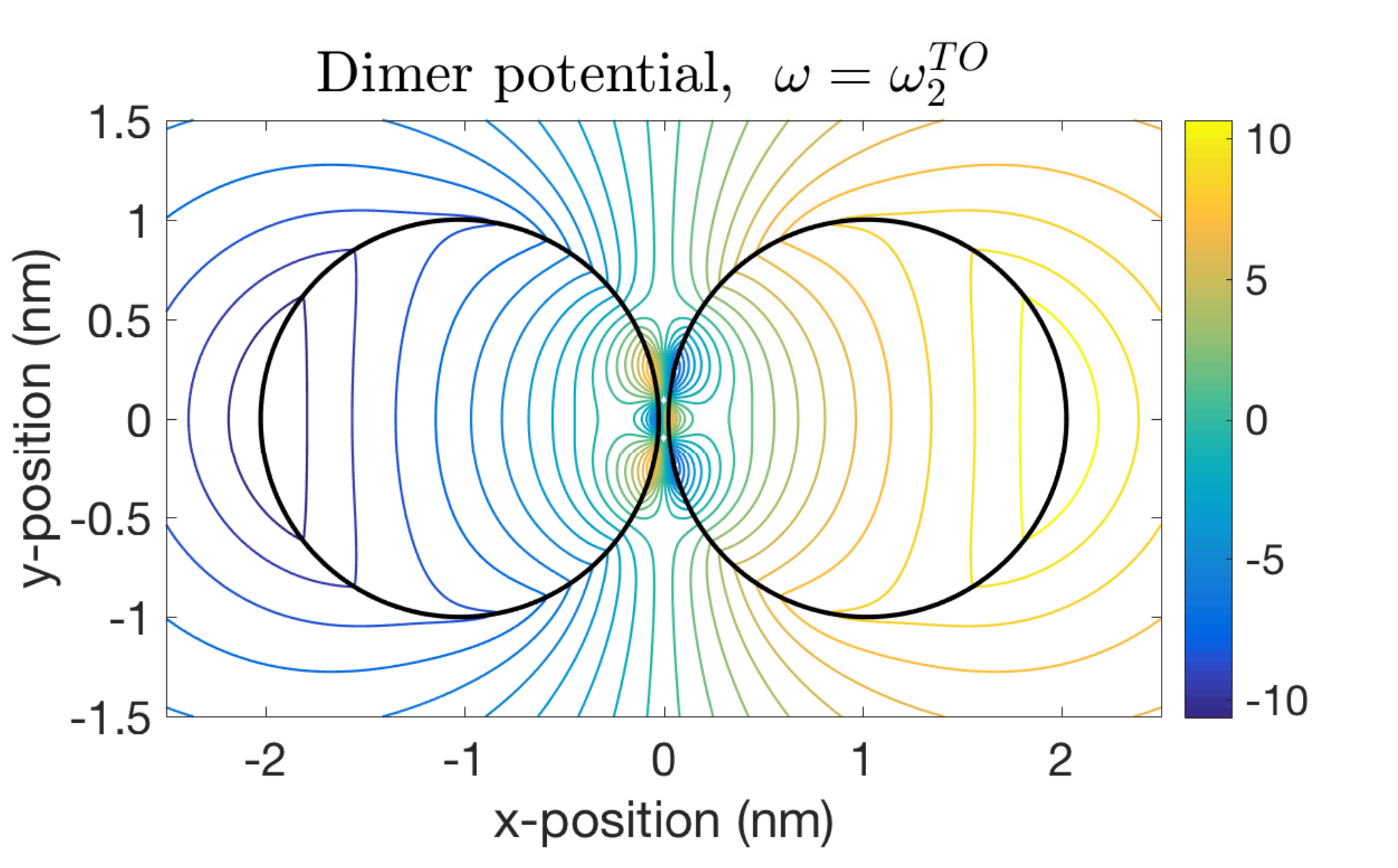}
\\[0.5em]
\includegraphics[width=8.5cm]{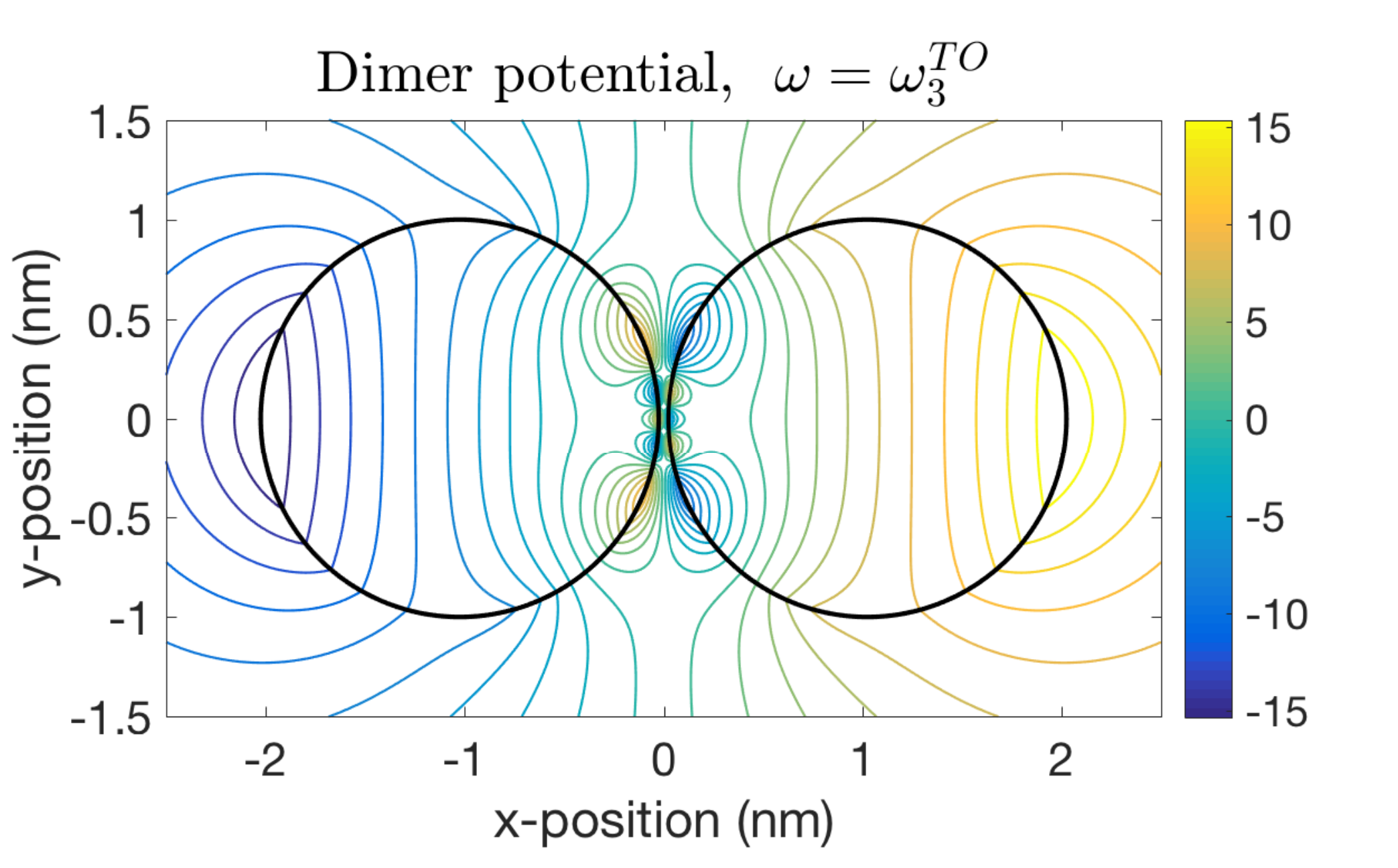}
\caption{
Potential distributions for the dimer. Assuming the $x$-polarized incident light, we consider the three cases when $\omega=\omega_1^{TO},\omega_2^{TO}$ and $\omega_3^{TO}$ so that the dimer plasmon potentials $V_1^{TO},V_2^{TO}$ and $V_3^{TO}$ are excited, respectively. We set $R=20$ nm, $\delta = 1$ nm, $\omega_p = 3.85$ eV and $\gamma = 0.1$ eV.
}
\label{figS:dimer}
\end{figure*}

We consider the Drude model $\epsilon = 1-\omega_p^2/(\omega(\omega+i\gamma))$ for the permittivity of the dimer $D=B_+\cup B_-$. We further assume that a uniform electric field is applied in the $x$-direction with  intensity $E_0$. Then the analytic expression for the total elctric potential $V$ is given  by \cite{Pendry_twoparticles1,PendryTO2012,PendryTO2013,PendryTO2015,PendryTO2017}:
\begin{equation}\label{dimer_potential}
V(\mathbf{r})=- E_0 x + \sum_{n=1}^\infty \frac{2 \omega_p^2 a e^{-2ns}}{\omega^2 - \omega_n^2 + i \gamma \omega} V_n^{TO}(\mathbf{r}),
\end{equation}
where $V_n^{TO}$ is defined by
\begin{equation}
V_n^{TO}(\mathbf{r})=V_n^{TO}(\Phi^{-1}(\mathbf{r}')) = 
\begin{cases}
e^{ns} \sinh (ns) e^{n \ln r'} \cos n\theta', &\quad \mathbf{r} \in B_-,
\\
 \sinh (n \ln r') \cos n\theta', &\quad \mathbf{r} \in \mathbb{R}^2\setminus(B_-\cup B_+),
\\
e^{ns} \sinh (ns) e^{-n \ln r'} \cos n\theta', &\quad \mathbf{r} \in B_+.
\end{cases}
\end{equation}
The function $V_n^{TO}$ represents the potential generated by the plasmon charge density $|\omega_n^{TO}(B_-,B_+)\rangle$ up to a multiplicative constant.
When $\omega$ is close to a resonance frequency $\omega_n$, the total potential $V$ is dominated by the corresponding dimer plasmon potential $V_n^{TO}$.
In Fig. \ref{figS:dimer}, we plot the imaginary part of the total potential $V$ when the frequency is $\omega=\omega_1^{TO}$, $\omega_2^{TO}$, and $\omega_3^{TO}$. 
We set $R=20$ nm and $\delta = 1$ nm. For the Drude model, we use the same values as in the previous section.

\section{Comparison between TO approach and the standard hybridization method}

Here we numerically compare the TO approach with the standard hybridization method by computing the electric field at the gap center (or the origin $(0,0)$) of the dimer $D=B_-\cup B_+$. Again, we assume that a uniform electric field is applied in the $x$-direction with  intensity $E_0$. In this case, due to the symmetry of the dimer, the electric field at the origin is aligned in the $x$-direction.
We also set the radius $R = 20$ nm and the gap distance $\delta = 0.1$ nm and use the Drude model with $\omega_p = 3.85$ eV and $\gamma = 0.1$ eV.

In the standard hybridization method,  one uses $\{ \cos m \theta_j, \sin m\theta_j\}_{m=1}^\infty$ as basis of the charge density $\sigma$ for each circular particle $B_j$. Here, $(r_j,\theta_j)$ represents the polar coordinates with respect to the center of $B_j$. For numerics, we truncate the basis and only consider  $\{ \cos m \theta_j, \sin m\theta_j\}_{m=1}^M$ for some positive integer $M$. 
 In fact, they are individual particle plasmons for the circular particle $B_j$. 
 Then the electric field can be computed by applying the standard multipole expansion technique.
In the TO approach, one can use Eq. \ref{dimer_potential} for computing the electric field. One should use the truncated TO basis $\{ |\omega_n^{TO}\rangle\}_{n=1}^N$ for a positive integer $N$.

\begin{figure*}
\centering
\includegraphics[width=6cm]{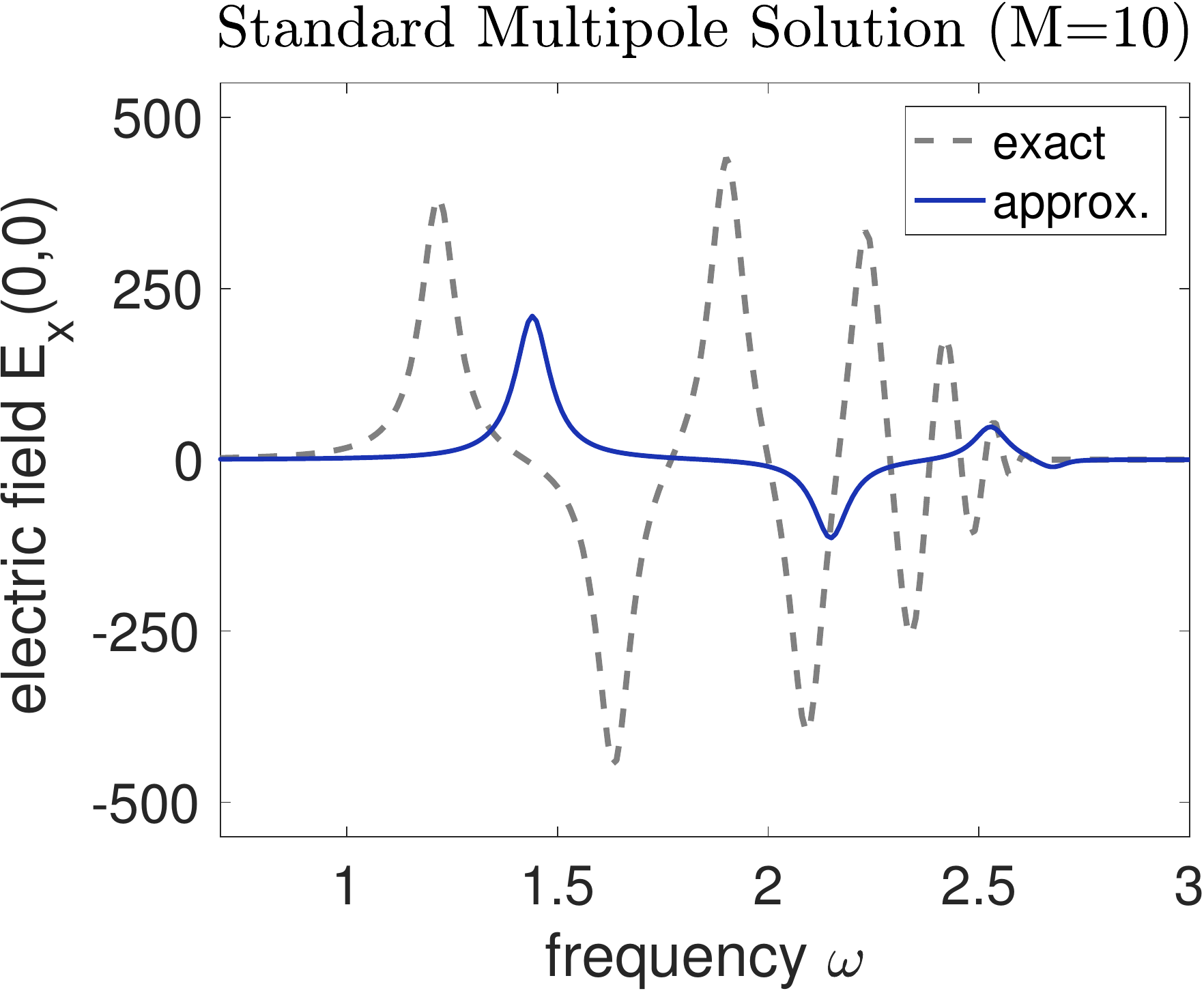}
\hskip.5cm
\includegraphics[width=6cm]{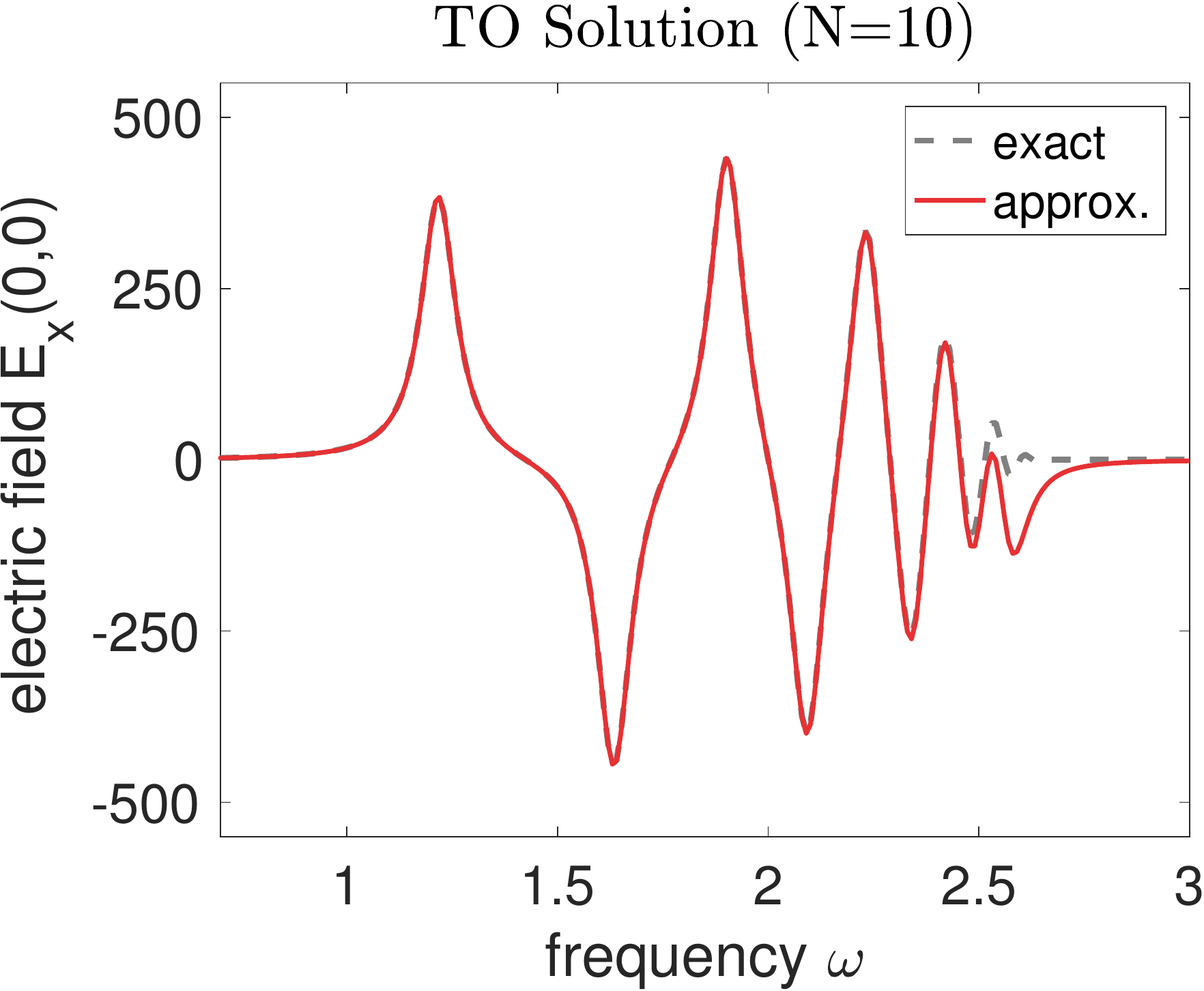}
\\[0.5em]
\includegraphics[width=6cm]{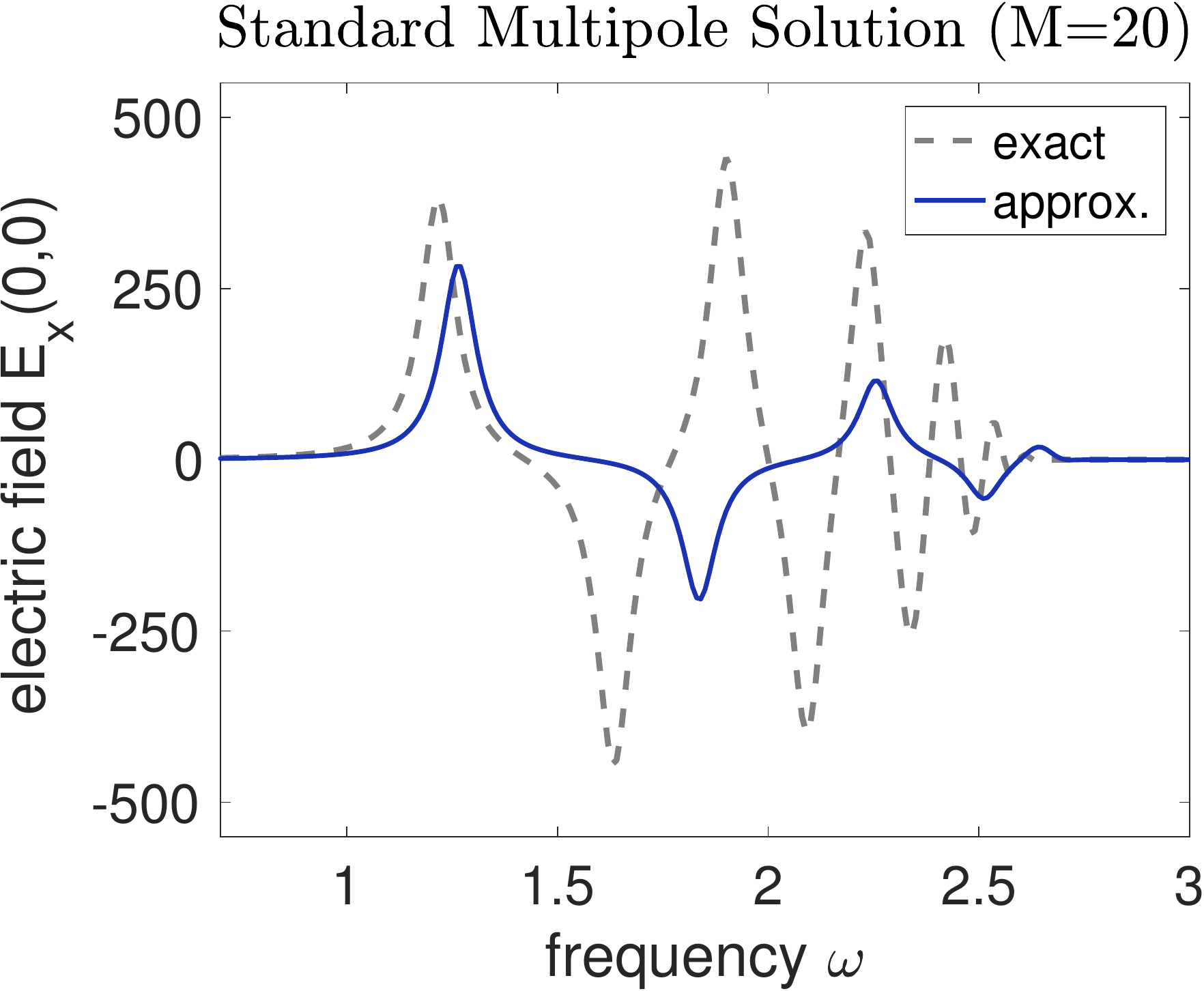}
\hskip.5cm
\includegraphics[width=6cm]{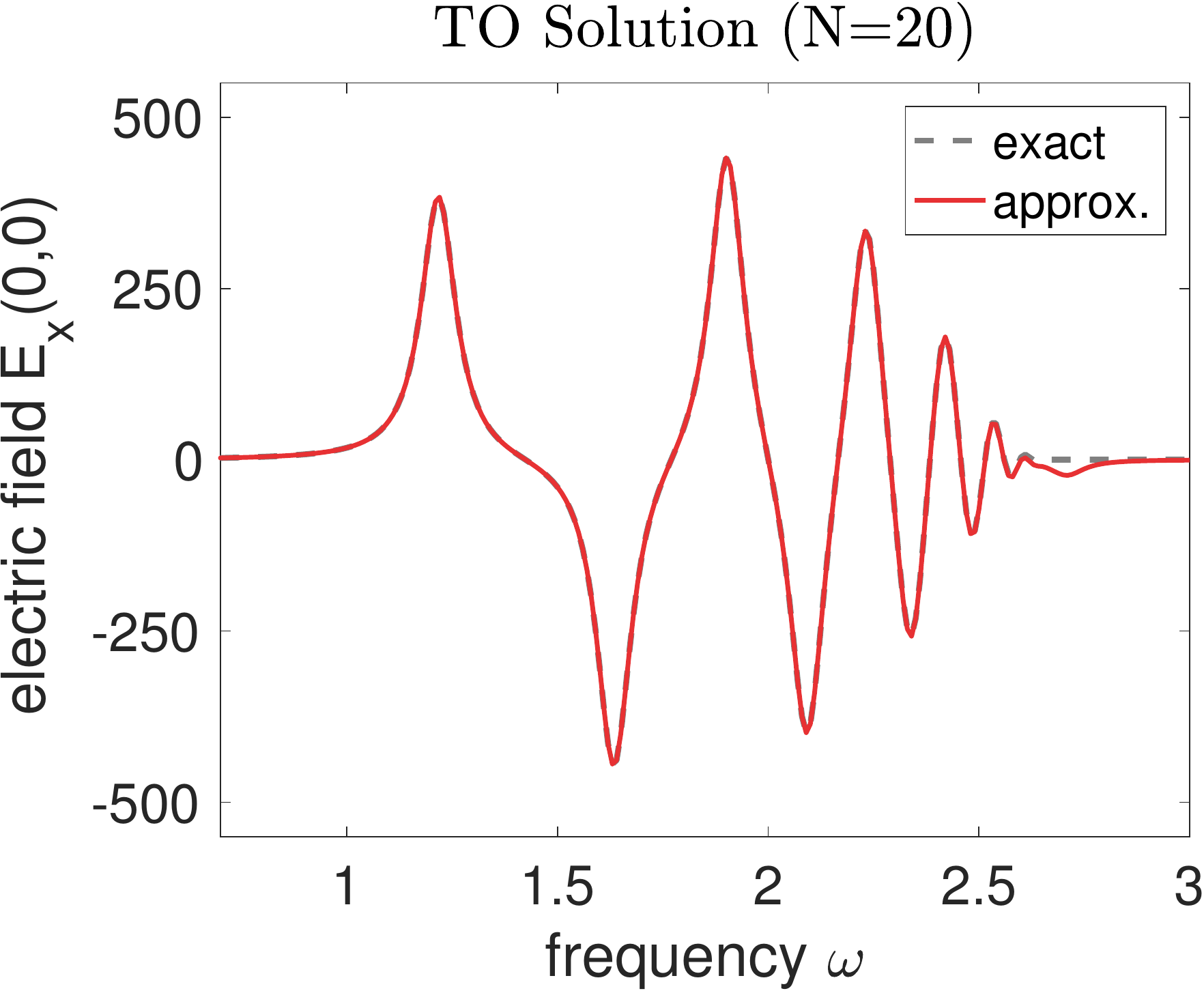}
\\[0.5em]
\includegraphics[width=6cm]{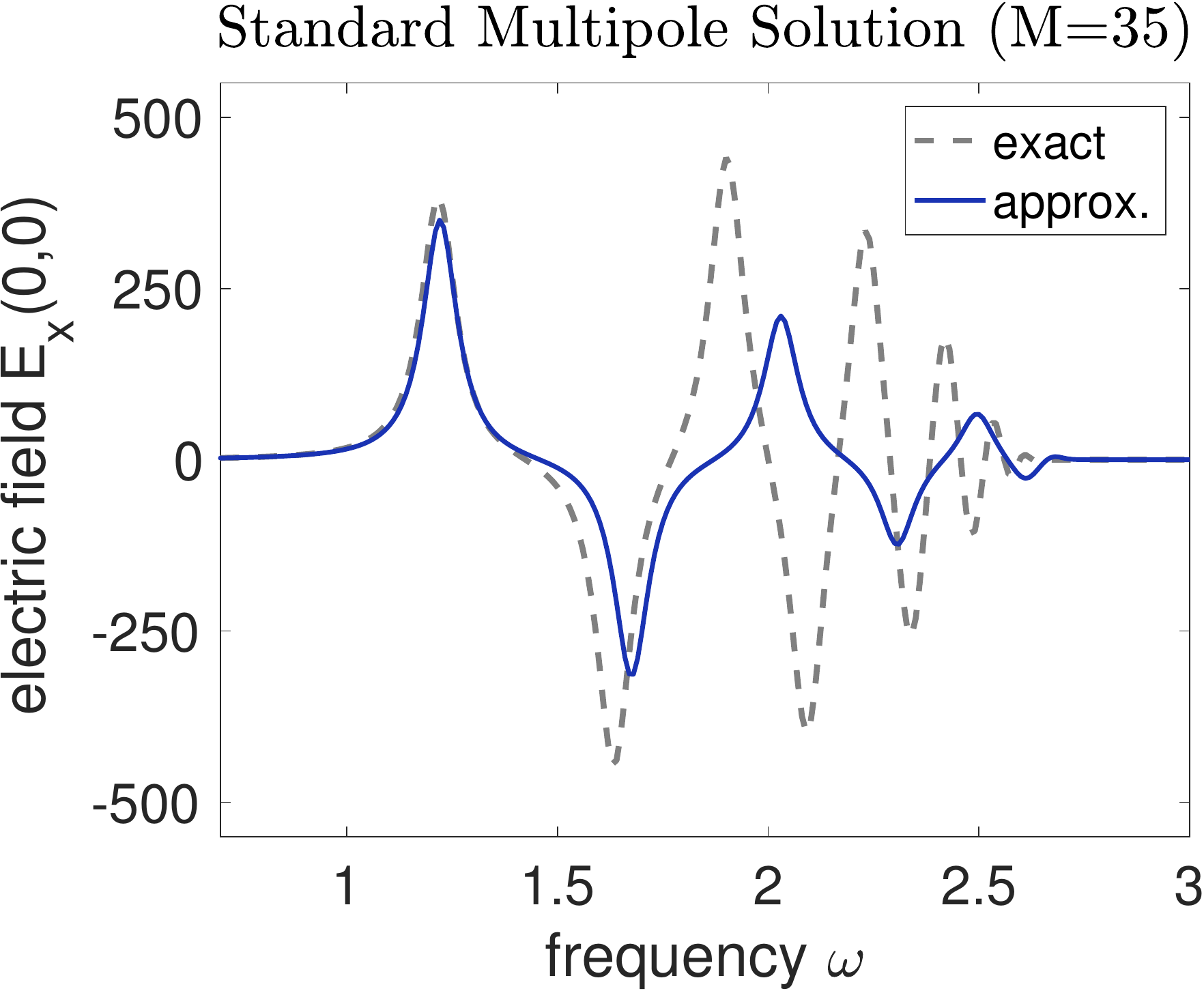}
\hskip.5cm
\includegraphics[width=6cm]{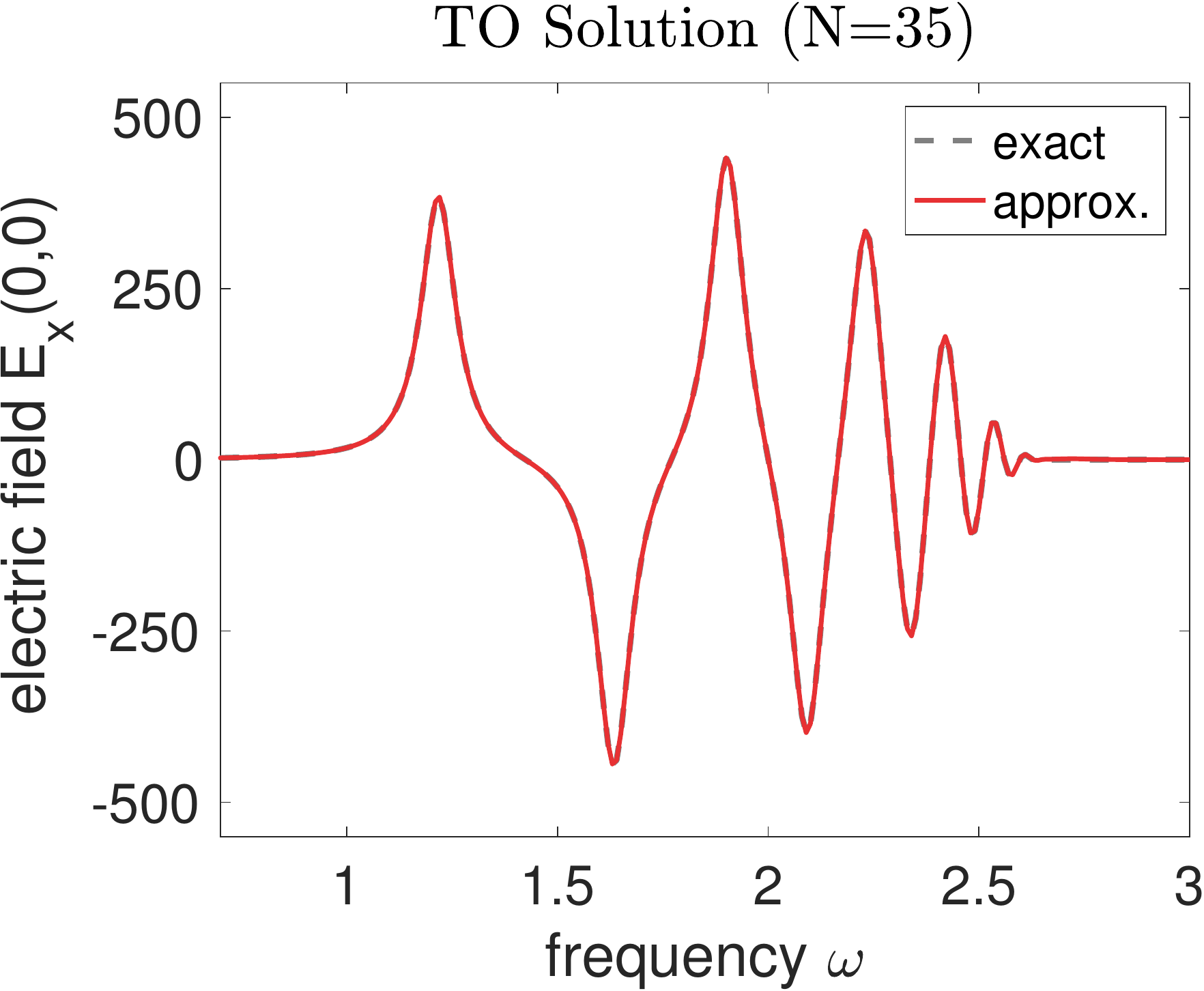}
\\[0.5em]
\includegraphics[width=6cm]{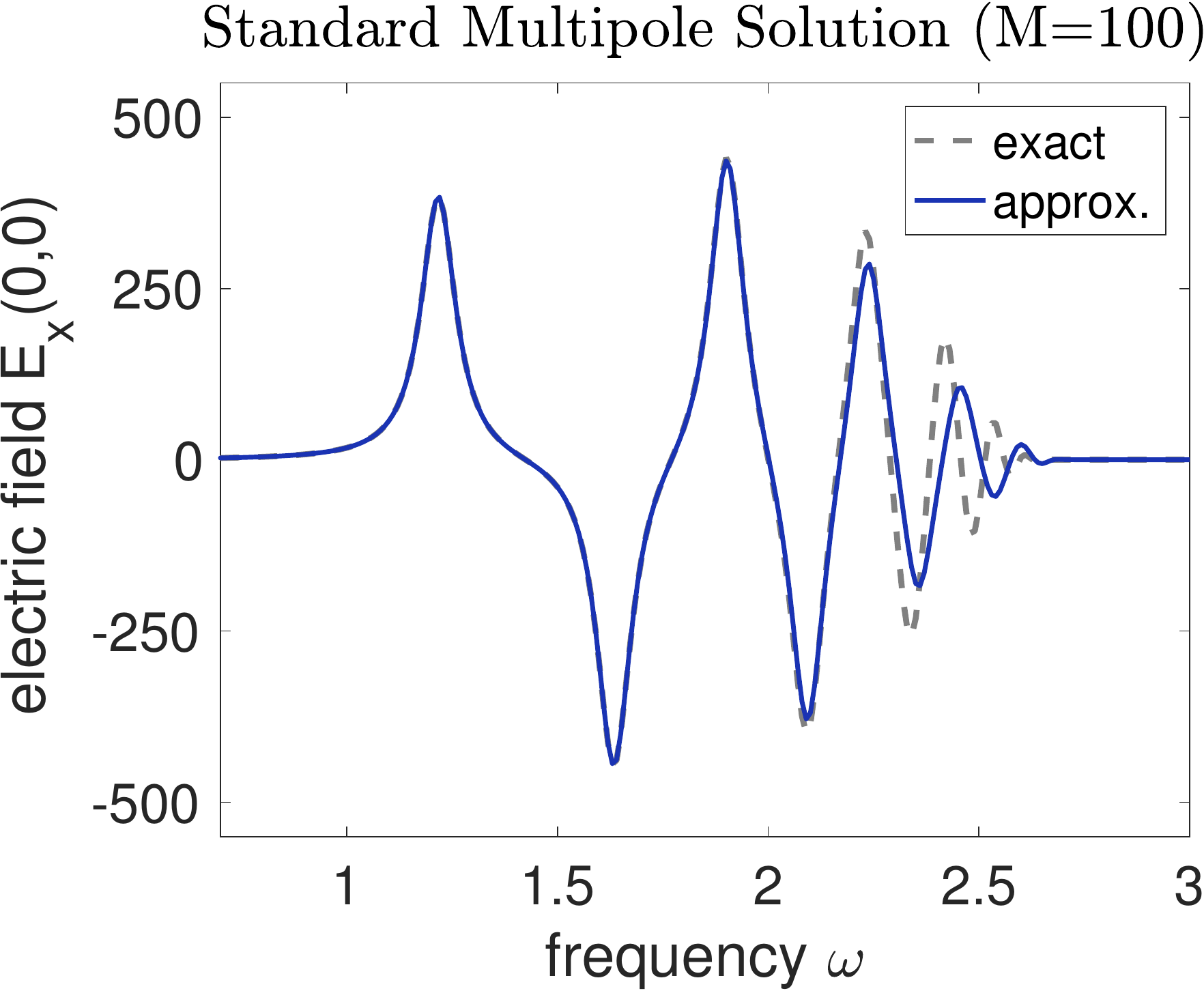}
\hskip.5cm
\includegraphics[width=6cm]{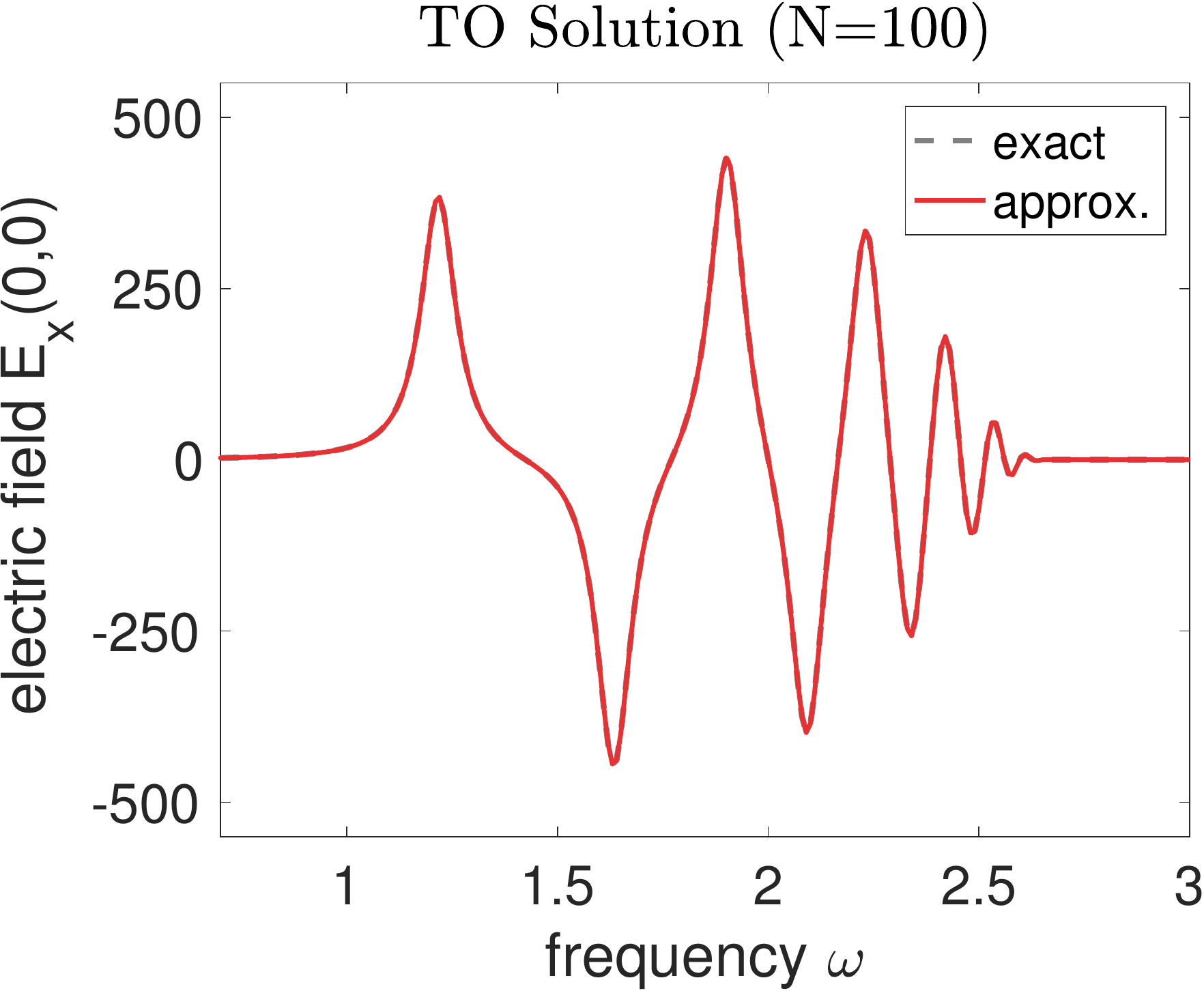}
\caption{
Comparison between TO approach and the standard hybridization method. (first column)
the electric field component $E_x(0,0)$ computed using the basis of standard hybridization method with the truncation order $M = 10, 20, 35, 100$. (second column) the electric field component $E_x(0,0)$ computed by the TO solution with the TO truncation order $N = 10, 20, 35, 100)$. 
We set  $R = 20$ nm, $\delta = 0.1$ nm, $\omega_p = 3.85$ eV and $\gamma = 0.1$ eV.
 }
 \label{figS:compare}
\end{figure*}

In Fig. \ref{figS:compare}, we plot the electric field component $E_x(0,0)$ as a function of the frequency.
In the first column (and the second column), the results are computed by the standard hybridization method (and the TO solution) when $M = 10, 20, 35, 100$ (and $N = 10, 20, 35, 100)$, respectively. 
The standard hybridization method gives an inaccurate result even when we use high number
of orders, say $M = 100$. On the contrary, the accuracy of the TO solution
is pretty good although we use a low order, say $N = 10$. This indicates that the TO approach is more accurate and efficient than the standard hybrid method when the gap distance $\delta$ is extremely small.

\section{Hybridization of singular plasmons: a trimer case}
Next, we consider the trimer $T=B_1\cup B_2 \cup B_3$ given in the main manuscript. Recall that the pairs $(B_1,B_2)$ and $(B_2,B_3)$ are  close-to-touching while $B_1$ and $B_3$ are well-separated.
After some translation and rotation and by abuse of notation, we can define the TO dimer plasmons for the pair $(B_1,B_2)$ and the pair $(B_2,B_3)$ as follows:
\begin{equation}
| \omega_{n}^{TO}(B_1,B_2)\rangle
=
\begin{cases}
|\omega_n^{TO}\rangle &\quad \mbox{on } \partial B_1 \cup \partial B_2,
\\
0 &\quad \mbox{on } \partial B_3,
\end{cases}
\end{equation}
and
\begin{equation}
| \omega_{n}^{TO}(B_2,B_3)\rangle
=
\begin{cases}
0 &\quad \mbox{on } \partial B_1, 
\\
|\omega_n^{TO}\rangle &\quad \mbox{on } \partial B_2 \cup \partial B_3.
\end{cases}
\end{equation}
These two dimer plasmons hybridize to form new plasmon modes.
We approximate a hybridized plasmon mode $|\omega_n\rangle$ as a linear combination $|\omega_n\rangle = a_n |\omega_n^{TO}(B_1,B_2)\rangle + b_n |\omega_n^{TO}(B_2,B_3)\rangle$. 
This is a good approximation when the gap distance $\delta$ is small. In fact, we can prove that the set of $|\omega_n^{TO}(B_i,B_j)\rangle$ form an `almost' orthogonal basis. More precisely,
as $\delta\rightarrow 0$,
\begin{equation}
\langle \omega_n^{TO}(B_1,B_2) | \omega_{n'}^{TO}(B_2,B_3)\rangle \approx 0  \quad \mbox{for all } n,n'=1,2,3,\cdots,  
\end{equation}
and consequently, 
\begin{equation}
\langle \omega_n | \omega_{n'} \rangle \approx 0  \quad \mbox{for } n\neq n'.
\end{equation}
Using the fact that $\mathcal{A}_D|\omega_n^{TO}\rangle=(\omega_n^{TO})^2|\omega_n^{TO}\rangle$, we can easily see that
\begin{equation}
\begin{bmatrix}
(\omega_n^{TO})^2 & \Delta_{n}
\\
\Delta_{n} & (\omega_n^{TO})^2 
\end{bmatrix}
\begin{bmatrix}
a_n
\\
b_n
\end{bmatrix}
=
\omega^2
\begin{bmatrix}
a_n
\\
b_n
\end{bmatrix},
\end{equation}
where $\Delta_n$ is given by
\begin{equation}\label{def_Deltan}
\Delta_n = \langle \omega_n^{TO}(B_1,B_2) |\mathcal{A}_T | \omega_n^{TO}(B_2,B_3) \rangle.
\end{equation}
By finding the eigenvalues and eigenvectors of the matrix on the LHS, we can find good approximations for the hybridized plasmons and their resonance frequencies
 as follows: 
\begin{equation}
|\omega_n^{\pm}  \rangle  \approx  \frac{1}{\sqrt{2}}\Big(|\omega_{n}^{TO}(B_1,B_2) \rangle \mp |\omega_{n}^{TO}(B_2,B_3) \rangle\Big), \quad n=1,2,\cdots,
\end{equation}
and their resonant frequencies are
\begin{equation}
\omega_n^{\pm} \approx \omega_n^{TO} \pm \Delta_n, \quad n=1,2,\cdots.
\end{equation}

So far, we have discussed the hybridization between the `anti-symmetric' dimer plasmons $|{\omega}_n^{TO}\rangle$.
Our SPH model can also describe the hybridization between `symmetric' dimer plasmons $|\widetilde{\omega}_n^{TO}\rangle$. By applying the same process, we obtain the another group of hybridized modes for the trimer as follows:
\begin{equation}\label{hybrid_trimer}
|\widetilde{\omega}_n^{\pm}  \rangle  \approx  \frac{1}{\sqrt{2}}\Big(|\widetilde{\omega}_{n}^{TO}(B_1,B_2) \rangle \pm |\widetilde{\omega}_{n}^{TO}(B_2,B_3) \rangle\Big), \quad n=1,2,\cdots,
\end{equation}
with their resonant frequencies 
\begin{equation}
\widetilde{\omega}_n^{\pm} \approx \widetilde{\omega}_n^{TO} \pm \widetilde{\Delta}_n, \quad n=1,2,\cdots.
\end{equation}
Here, $\widetilde{\Delta}_n$ is given by $\widetilde{\Delta}_n = -\langle \widetilde{\omega}_n^{TO}(B_1,B_2) |\mathcal{A}_T | \widetilde{\omega}_n^{TO}(B_2,B_3) \rangle \geq 0$.

By including a full set of basis (the gap-plasmons with different TO angular momenta $n$ and the gap-plasmons for the other pair $(B_1,B_3)$), we can compute all the resonant frequencies and their associated plasmon modes accurately. 
We emphasize that all the matrix elements, including $\Delta_n$ and $\widetilde{\Delta}_n$, for the operator $\mathcal{A}_T$ can be computed analytically by using the explicit formulas derived in \cite{sirev}, specifically, a 2D version of Theorem K.3 in Ref.\cite{sirev}.
The absorption cross section can be computed by using Eqs. \ref{spectral_decomp} and \ref{absor}.
As mentioned in the main manuscript,
it is straightforward to extend the above coupled mode theory to a more general system of particles.

We remark that our SPH model is also numerically efficient in the nearly touching case since the TO gap-plasmons capture the close-to-touching interaction analytically. 
In fact, the authors developed a numerical method for a similar problem in Ref. \cite{sirev}. This method can also deal with the close-to-touching particles system efficiently. But we emphasize that the SPH model is quite different and has several advantages. 
Firstly, contrary to the method in \cite{sirev}, the SPH model enables us to compute the resonant frequencies and their associated plasmon modes directly, providing a deep physical understanding of singular plasmons.
 Secondly, as the frequency $\omega$ changes, the method in \cite{sirev} requires to do the numerical inversion repeatedly, while the SPH model is free from this drawback. In view of the spectral decomposition formula in Eq. \ref{spectral_decomp} for the charge density $\sigma$, we only need to compute the eigenvalues and eigenfunctions of the NP operator $\mathcal{K}_\Omega^*$, which is independent of the frequency $\omega$, only once. 
 Roughly speaking, the aforementioned two drawbacks of the previous method originate from the use of the image charge series, which depends on the permittivity (hence the frequency through the Drude model), for capturing the close-to-touching interaction.

\section{Potential distributions of the trimer plasmons}

To illustrate the hybridization of gap-plasmons clearly, we plot the potential distributions at the trimer resonance frequencies $\omega = \omega_n^\pm$.
We assume the radius $R$ of each particle is $R=20$ nm. For the permittivity of the particles, we use the Drude model $\epsilon = 1-\omega_p^2/(\omega(\omega+i\gamma))$ with $\omega_p = 3.85$ eV and $\gamma = 0.1$ eV.

\begin{figure}
\centering
\includegraphics[width=7.5cm]{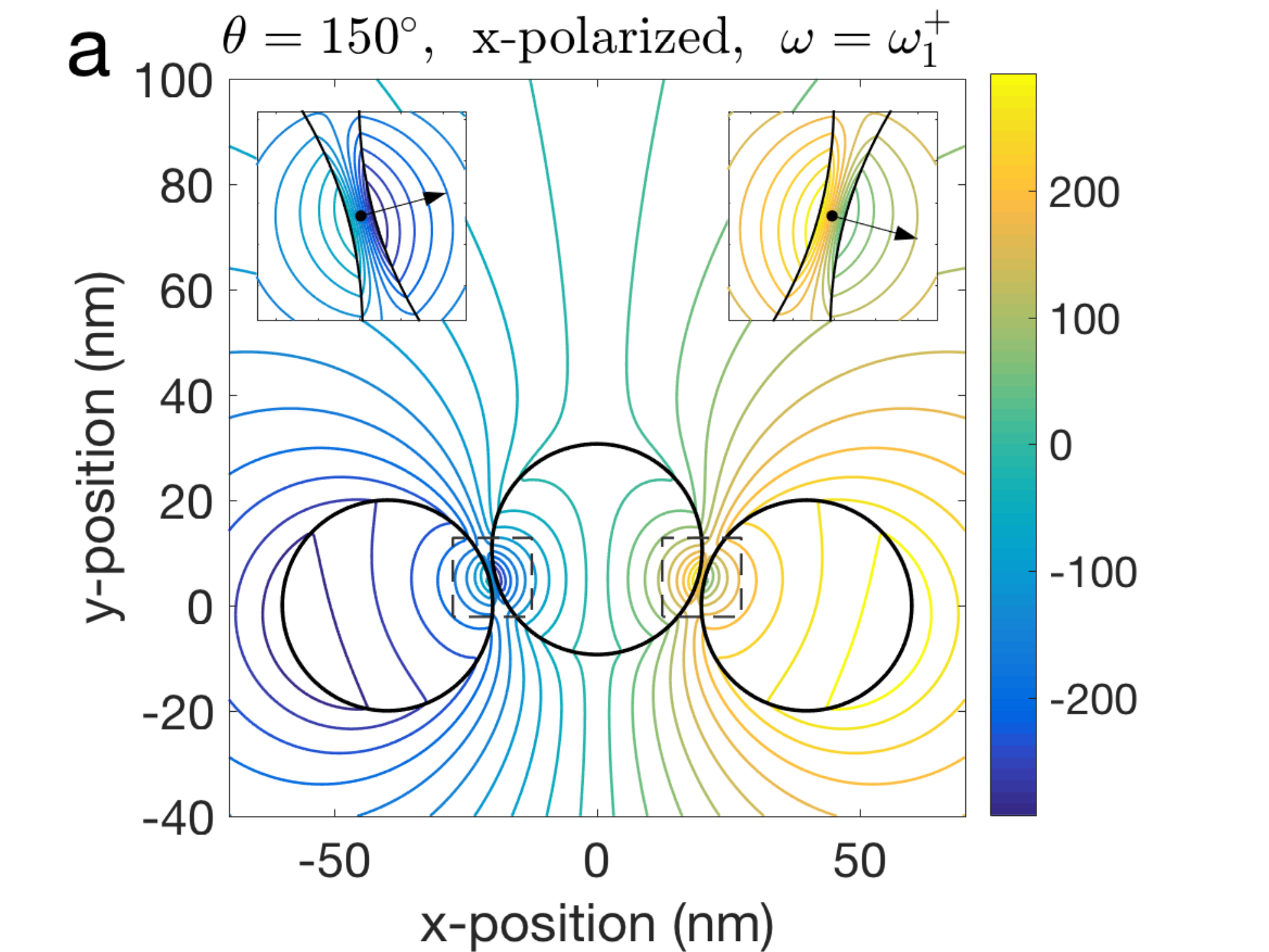}
\hskip-.5cm
%\\[0.75em]
\includegraphics[width=7.5cm]{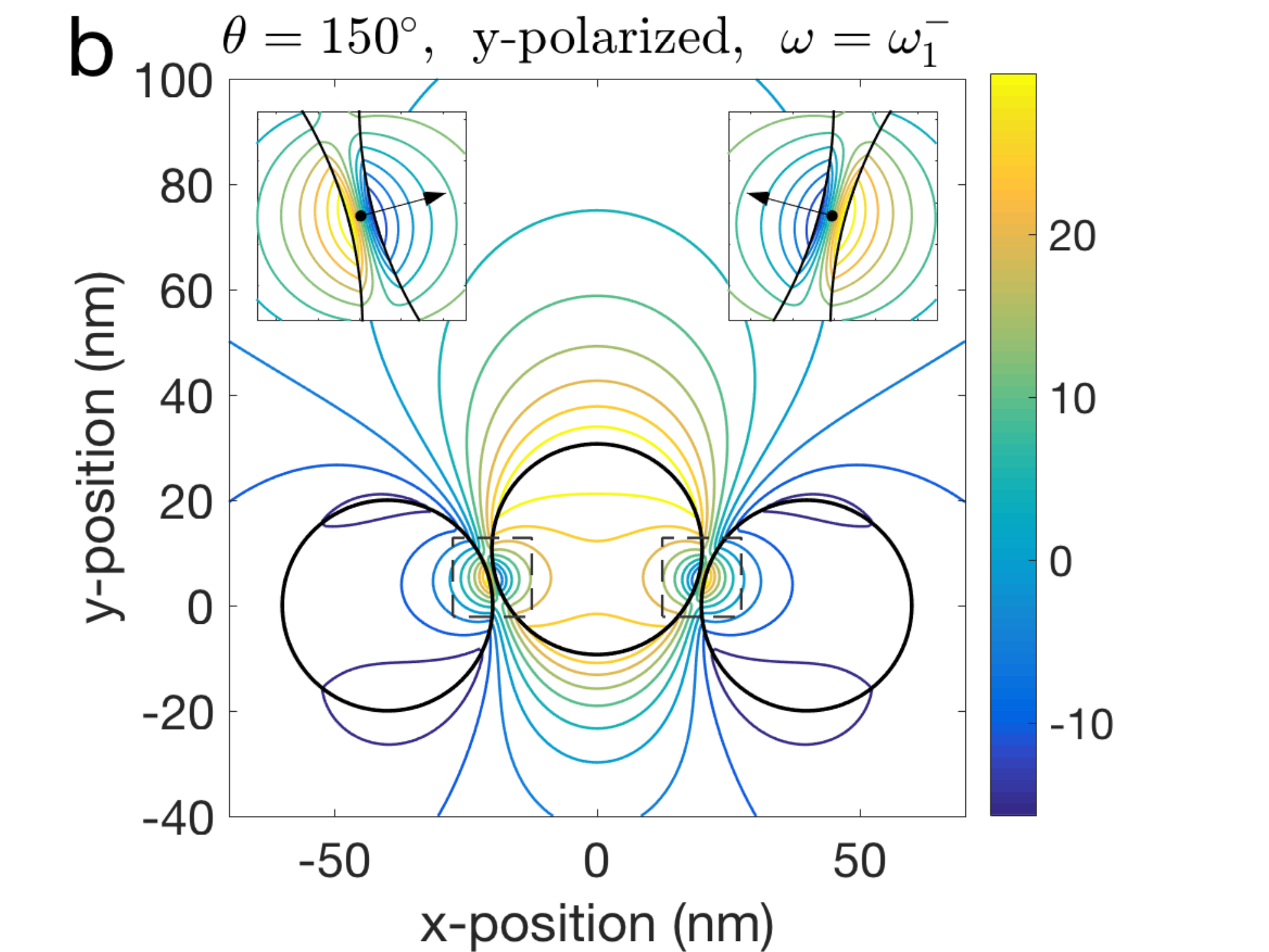}
\caption{
Imaginary parts of the electric potentials for the trimer with the gap distance $\delta = 1.25$ nm and the bonding angle $150^\circ$ when (A) the frequency $\omega$ is $\omega = \omega_1^+$ and the $x$-polarized light is incident; (B) the frequency $\omega$ is $\omega = \omega_1^-$ and the $y$-polarized light is incident. (small insets) Zoomed plots on the gap regions for the pairs $(B_1, B_2)$ and $(B_2,B_3)$. The black arrows mean the electric fields at the gap centers. We set $R = 20$ nm, $\omega_p = 3.85$ eV and $\gamma = 0.1$ eV.
}
\label{figS:contour1}
\end{figure}

\begin{figure}
\centering
%\hskip-.9cm
\includegraphics[width=7.5cm]{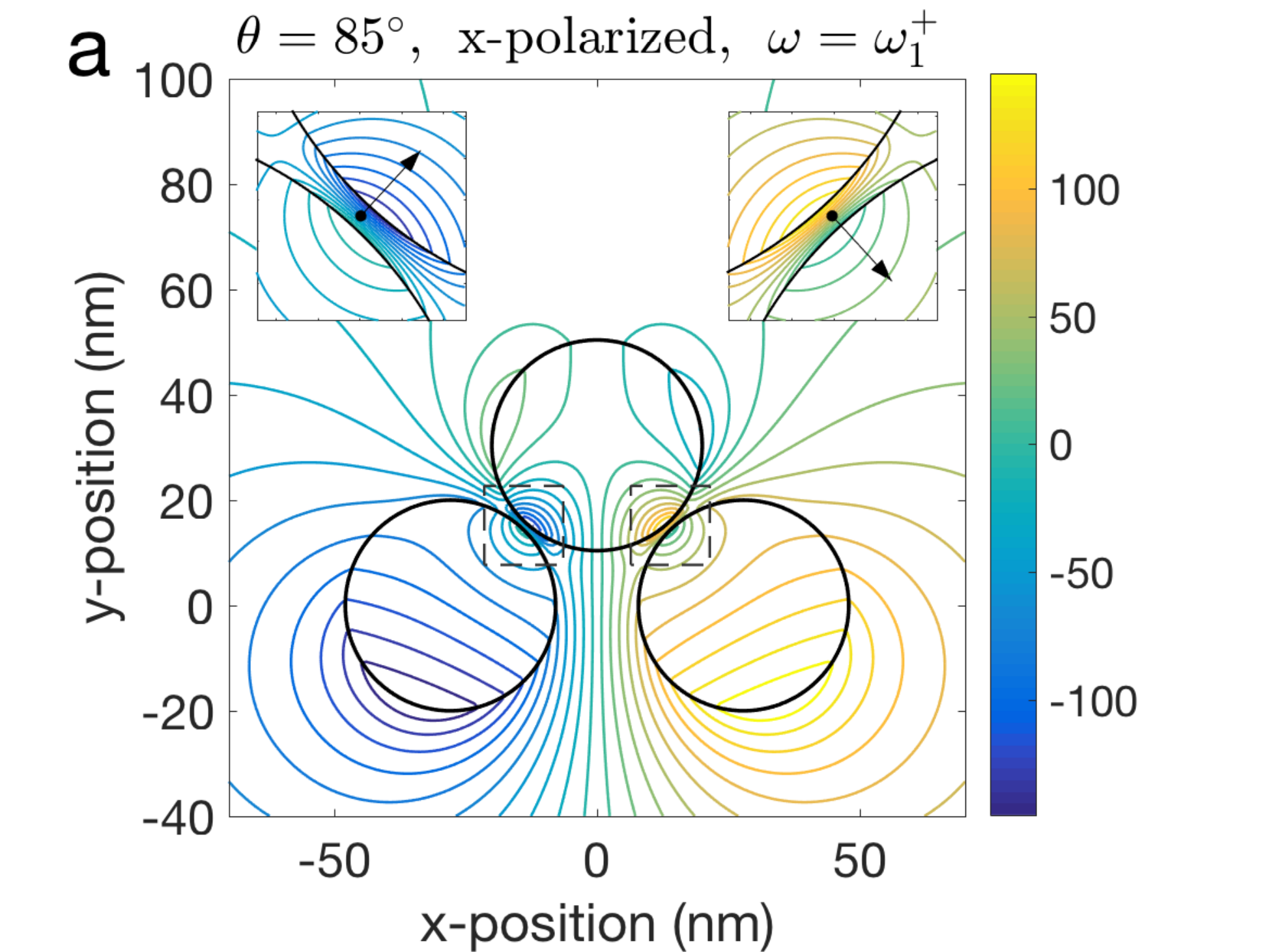}
%\\[0.75em]
\hskip-.5cm
\includegraphics[width=7.5cm]{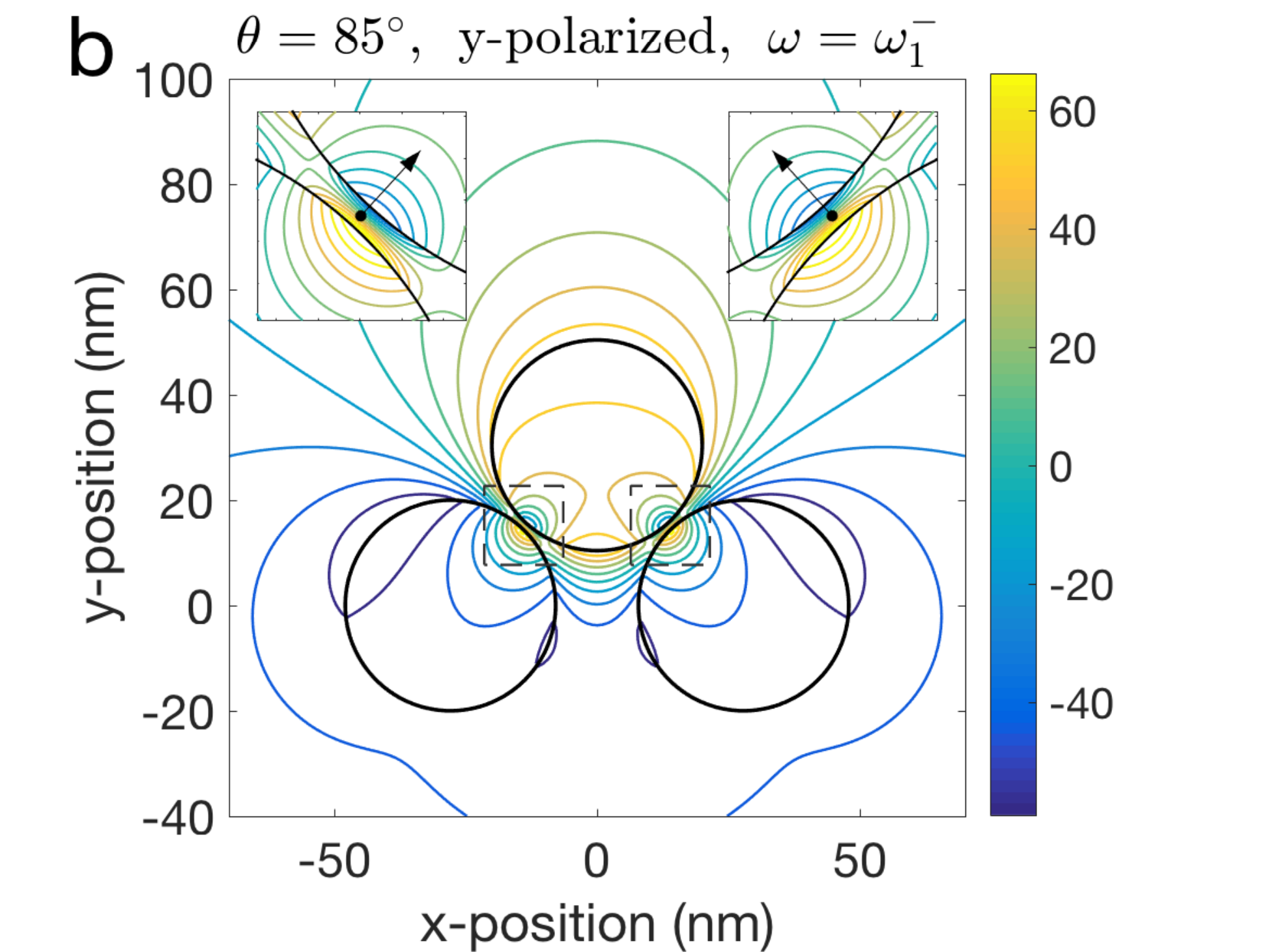}
\caption{
Imaginary parts of the electric potentials for the trimer with the gap distance $\delta = 1.25$ nm and the bonding angle $85^\circ$ when (A) the frequency $\omega$ is set as$\omega = \omega_1^+$ and the $x$-polarized light is incident; (B) the frequency $\omega$ is $\omega = \omega_1^-$ and the $y$-polarized light is incident. (small insets) Zoomed plots on the gap regions for the pairs $(B_1, B_2)$ and $(B_2,B_3)$. The black arrows mean the electric fields at the gap centers. We set $R = 20$ nm, $\omega_p = 3.85$ eV and $\gamma = 0.1$ eV.
}
\label{figS:contour2}
\end{figure}

In Fig. \ref{figS:contour1}, assuming the gap distance $\delta = 1.25$ nm and the bonding angle $\theta = 150^\circ$, we plot the imaginary part of the total potential $V$. We consider both cases of the $x$-polarized (Fig. \ref{figS:contour1}A) and $y$-polarized (Fig. \ref{figS:contour1}B) incident light.
We set the frequency $\omega$ as $\omega=\omega_1^+$ (and $\omega=\omega_1^-$) in the former (and the latter) case so that the trimer plasmon $|\omega_1^+\rangle$ (and $|\omega_1^-\rangle$) can be excited. In Fig. \ref{figS:contour1}A (or Fig. \ref{figS:contour1}B), it is clearly shown that the potential is dominated by an anti-bonding (or bonding) combination of the two TO dimer plasmon potentials for $|\omega_1^{TO}(B_1,B_2)\rangle  $ and $|\omega_1^{TO}(B_2,B_3)\rangle$ (see Fig. \ref{figS:dimer} for the dimer plasmon potentials). So they are in accordance with the theoretical prediction Eq. \ref{hybrid_trimer}. 
In Fig. \ref{figS:contour2}, we consider the another case with the same gap distance $\delta=1.25$ nm but with a smaller bonding angle $\theta = 85^\circ$.

\begin{figure*}
\centering
\includegraphics[width=7.5cm]{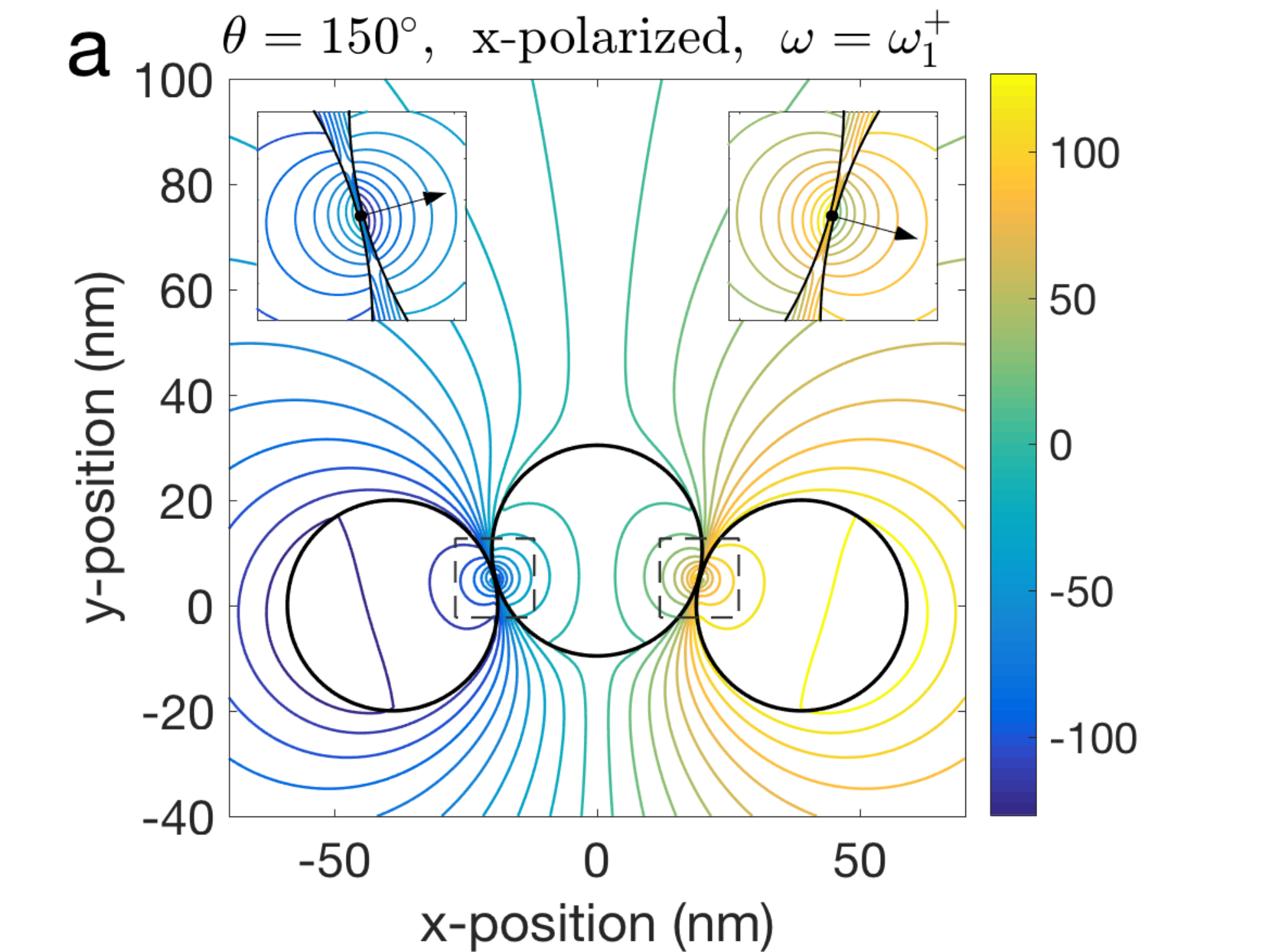}
\hskip-.5cm
\includegraphics[width=7.5cm]{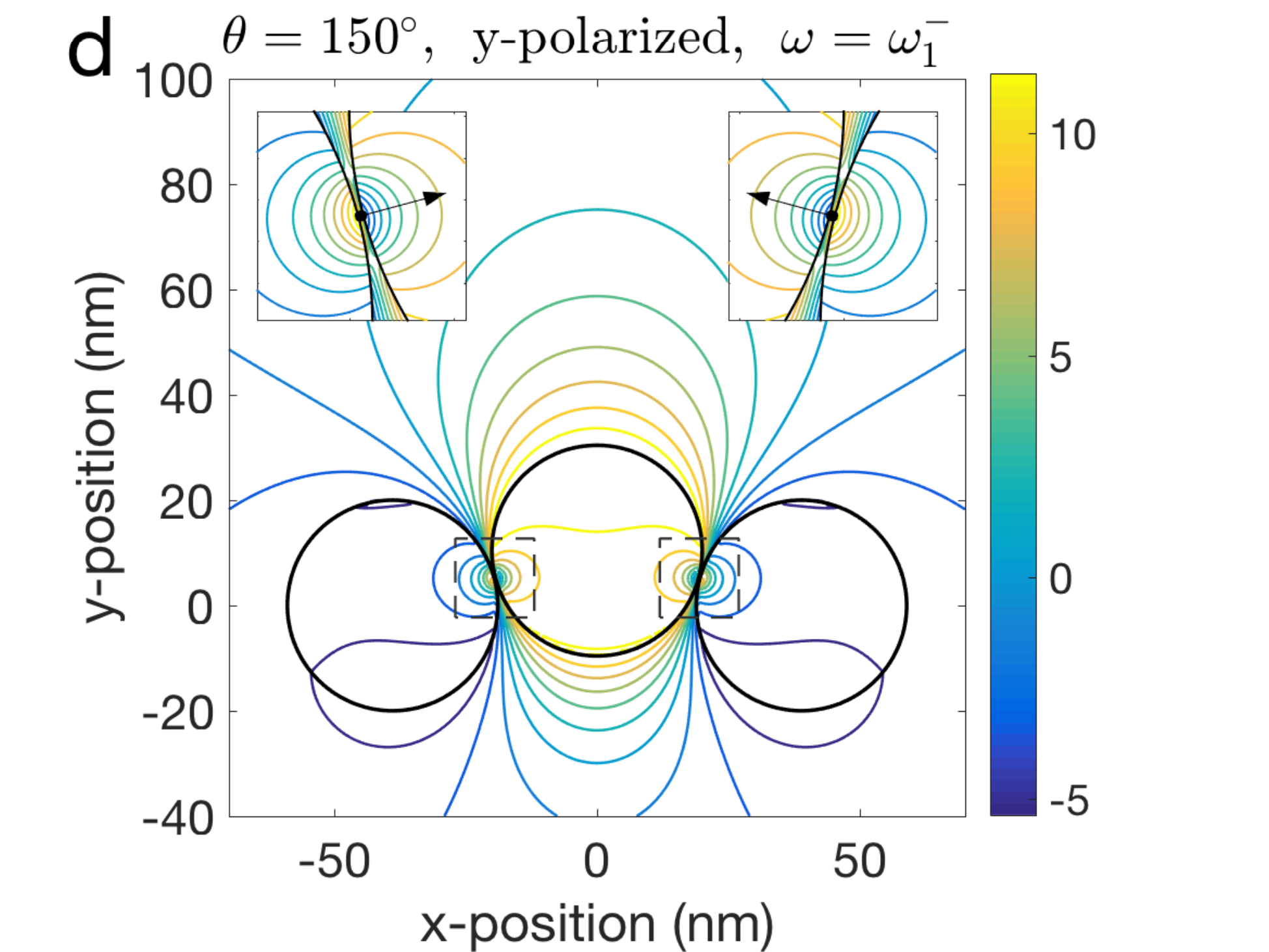}
\\[0.5em]
\includegraphics[width=7.5cm]{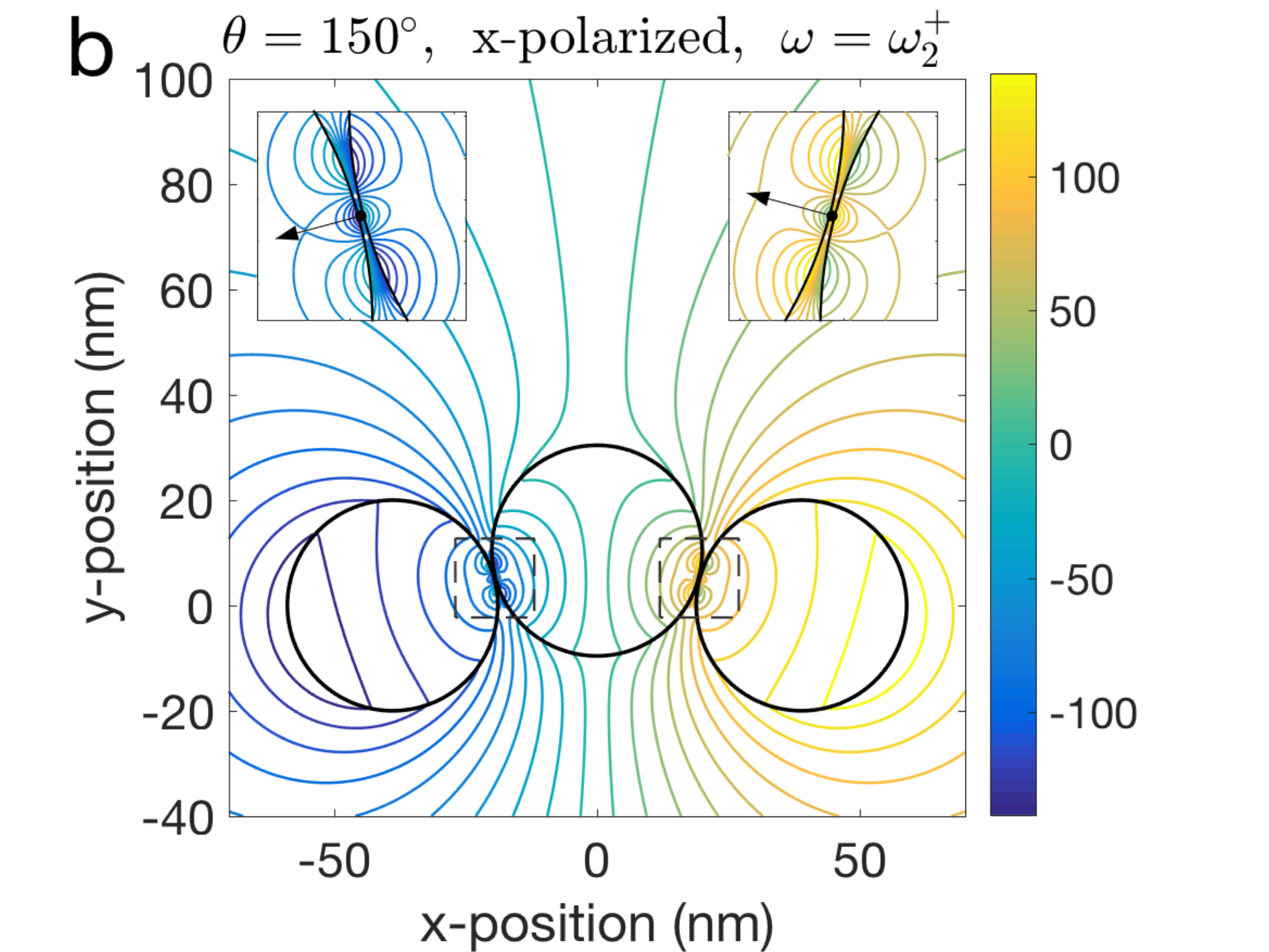}
\hskip-.5cm
%\\[0.75em]
\includegraphics[width=7.5cm]{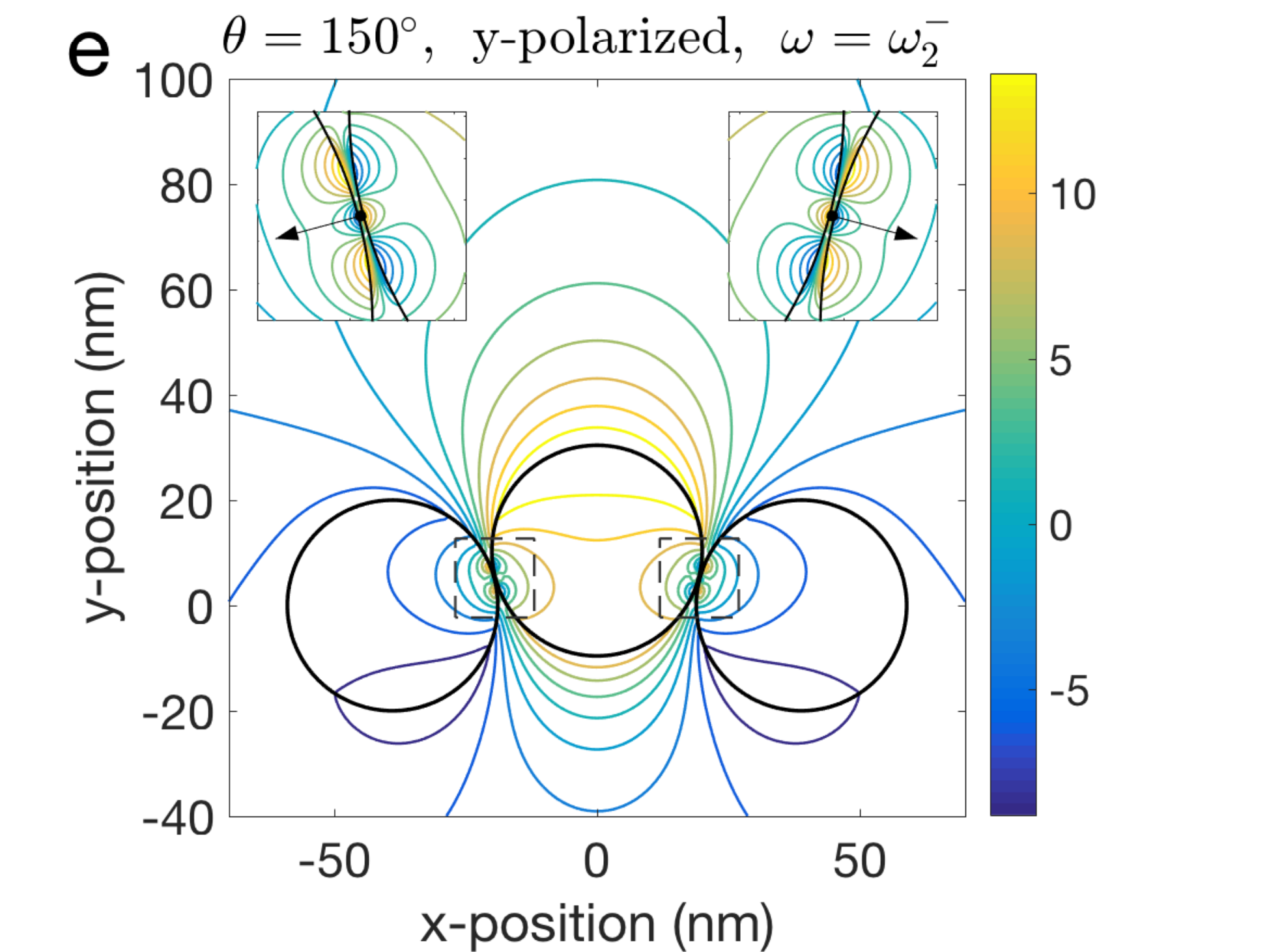}
\\[0.5em]
\includegraphics[width=7.5cm]{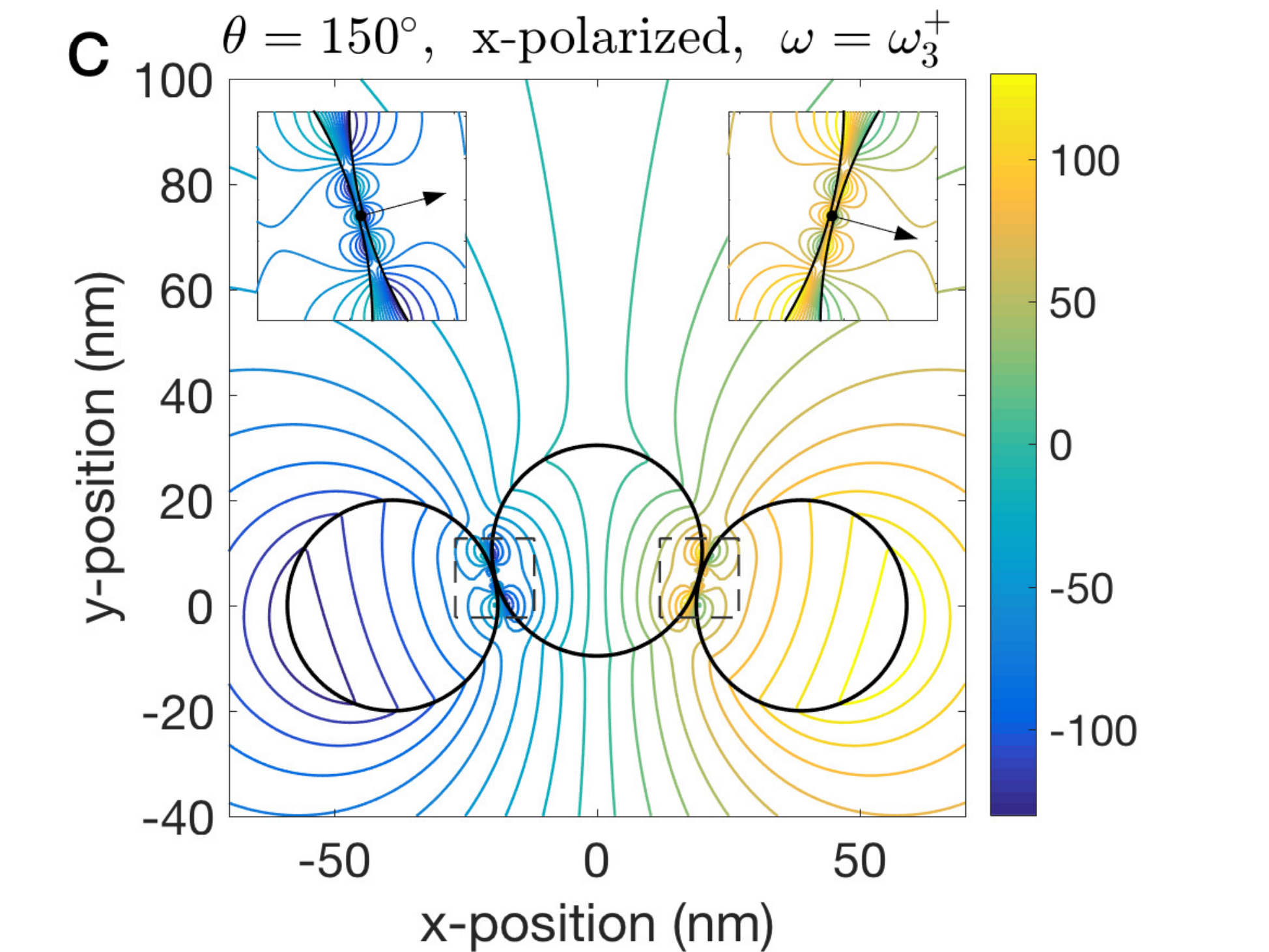}
\hskip-.5cm
\includegraphics[width=7.5cm]{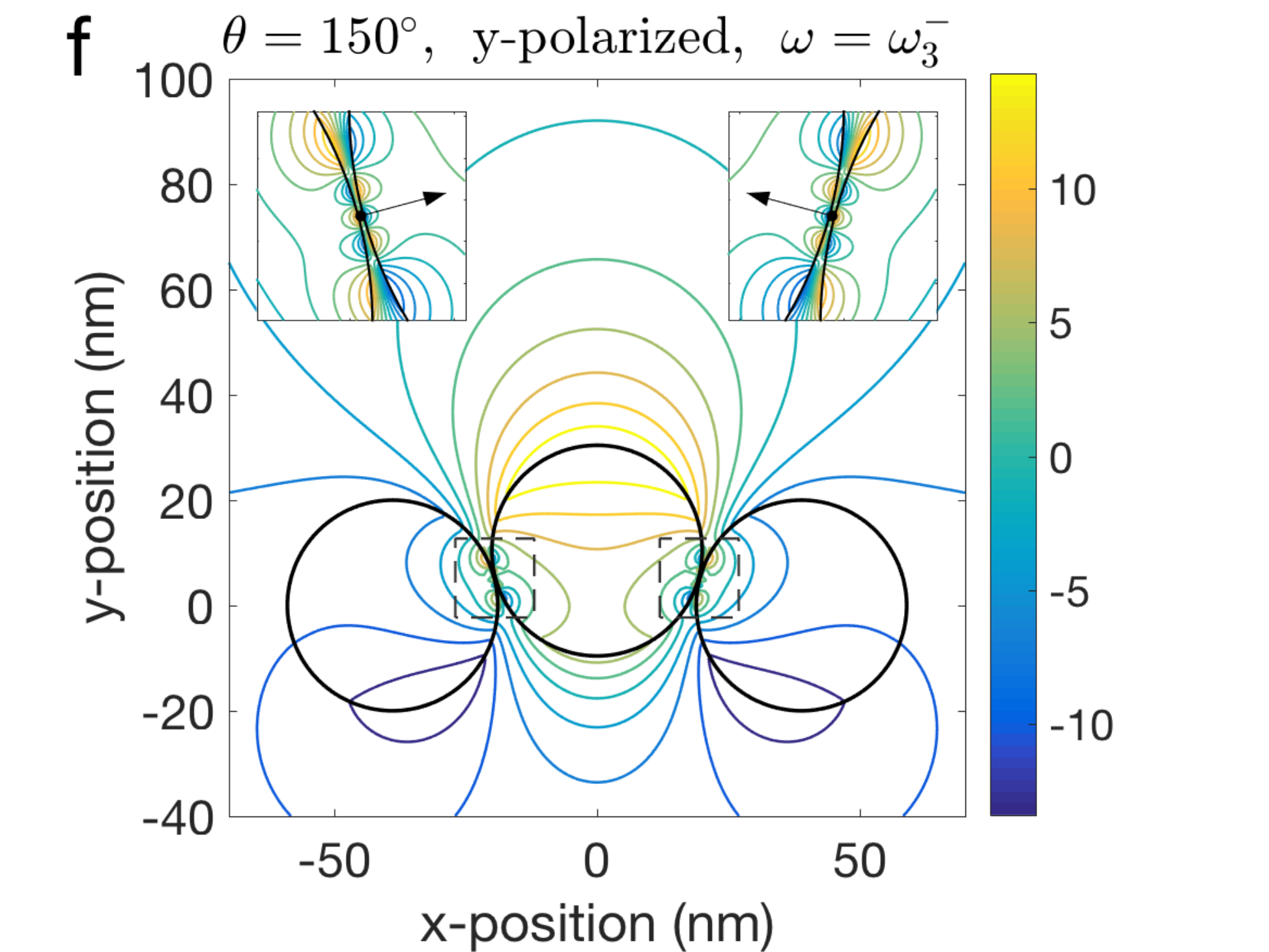}
\caption{
Imaginary parts of the electric potentials for the trimer with the gap distance $\delta = 0.25$ nm and the bonding angle $150^\circ$ when (A,B,C) the frequency $\omega$ is set as $\omega = \omega_1^+, \omega_2^+, \omega_3^+$ and the $x$-polarized light is incident; (B) the frequency $\omega$ is $\omega = \omega_1^-, \omega_2^-, \omega_3^-$ and the $y$-polarized light is incident. (small insets) Zoomed plots on the gap regions for the pairs $(B_1, B_2)$ and $(B_2,B_3)$. The black arrows mean the electric fields at the gap centers. We set $R = 20$ nm, $\omega_p = 3.85$ eV and $\gamma = 0.1$ eV.
}
\label{figS:contour3}
\end{figure*}

\begin{figure*}
\centering
\includegraphics[width=7.5cm]{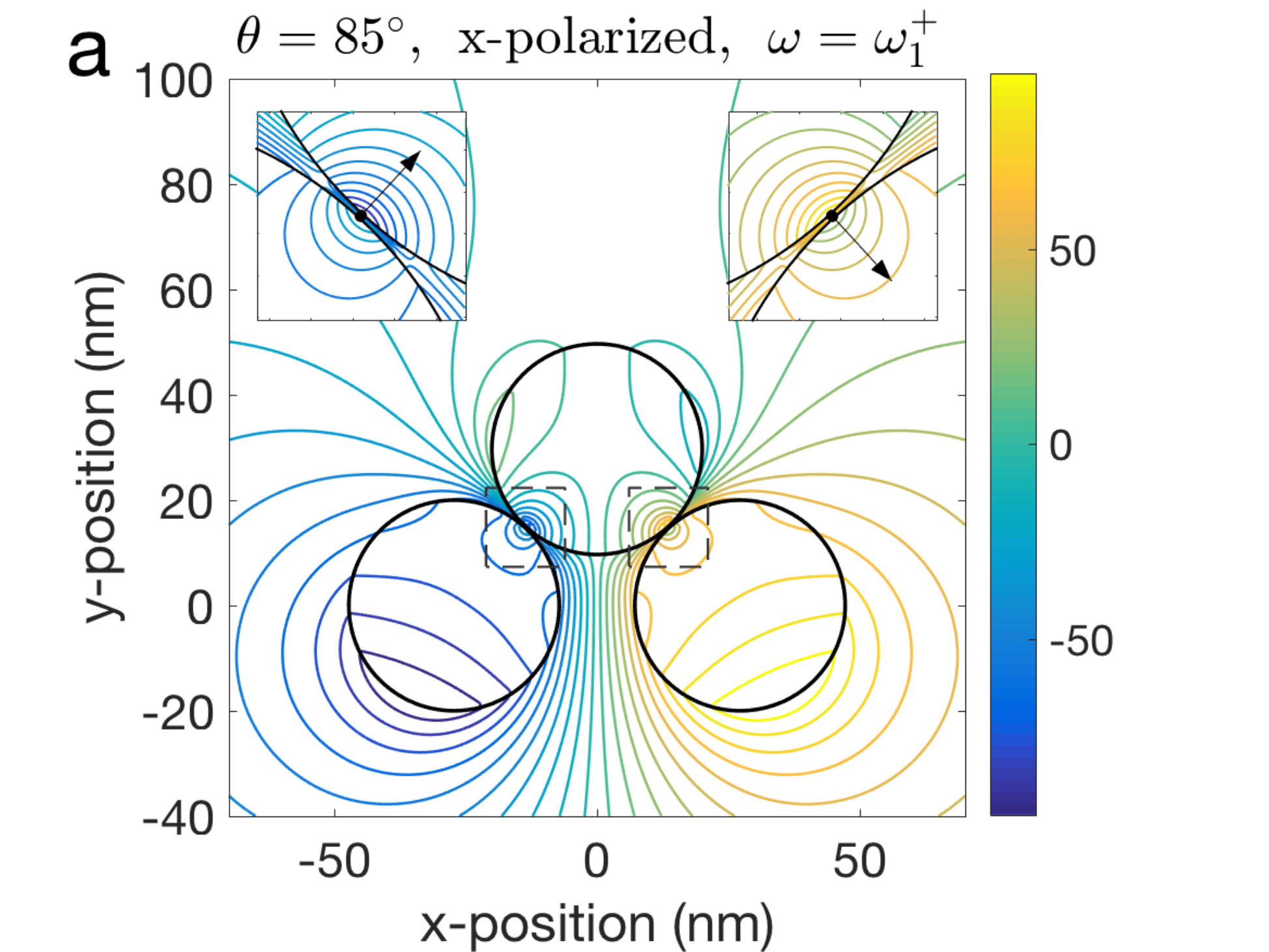}
\hskip-.5cm
\includegraphics[width=7.5cm]{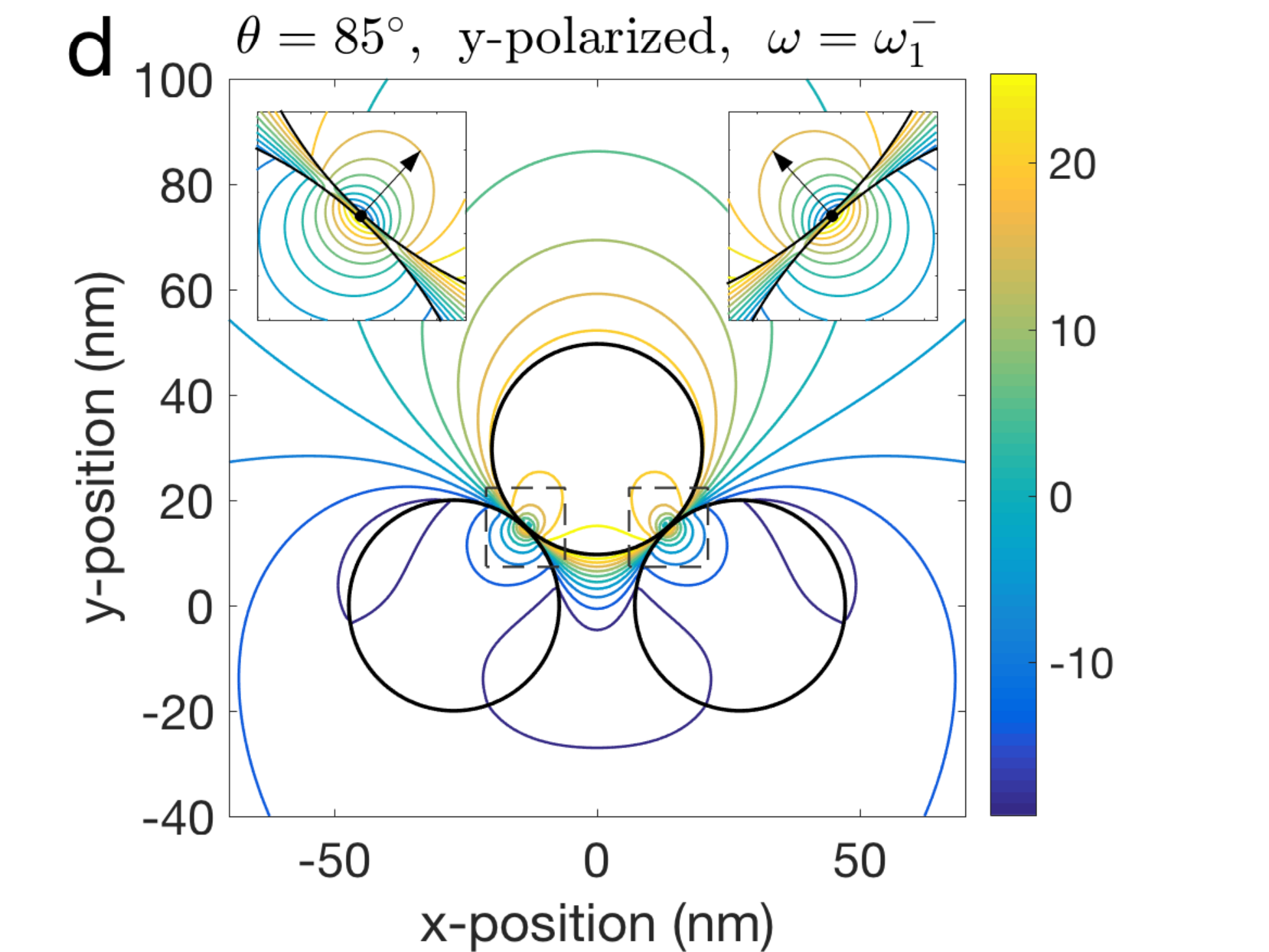}
\\[0.5em]
\includegraphics[width=7.5cm]{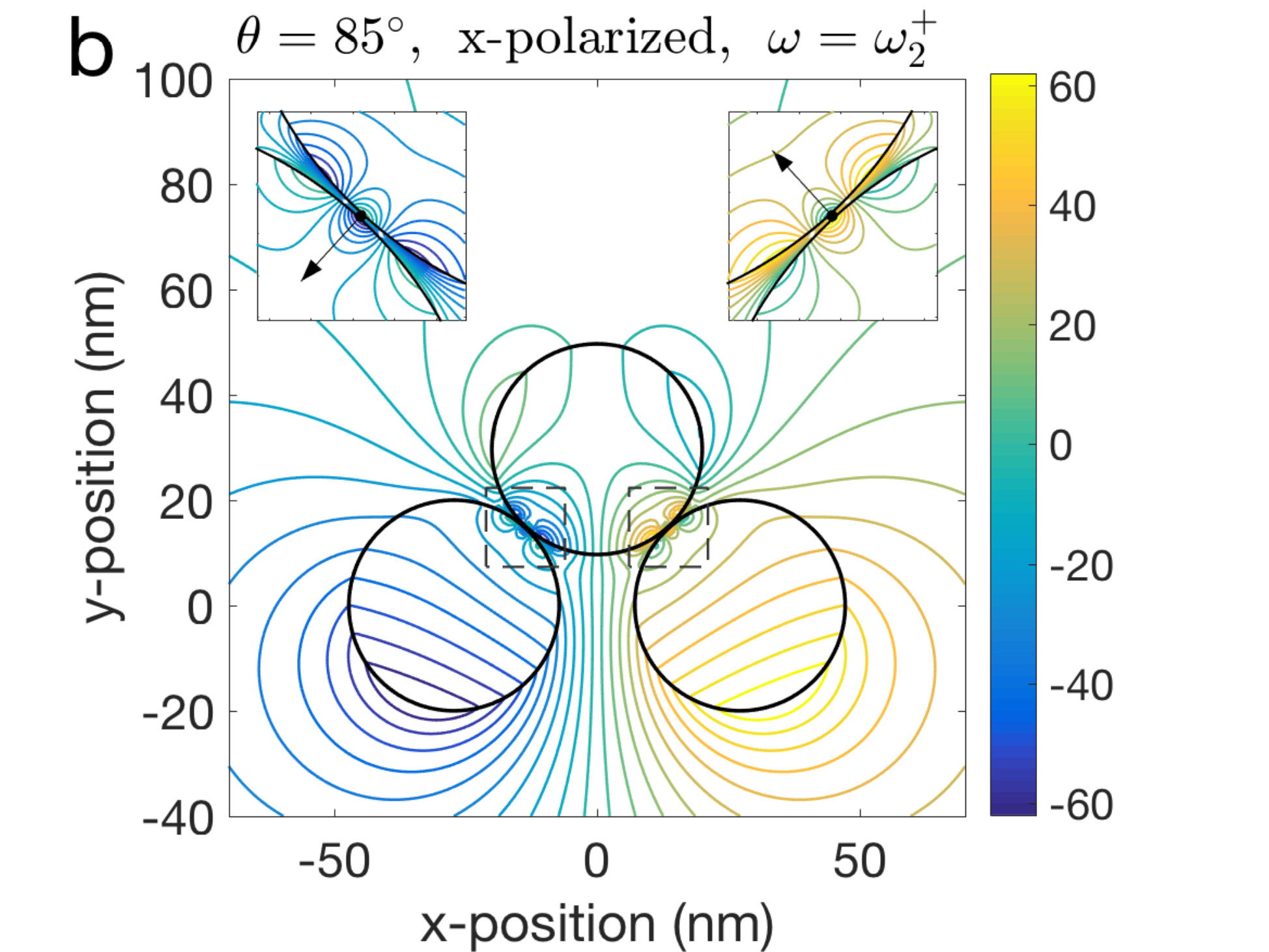}
\hskip-.5cm
\includegraphics[width=7.5cm]{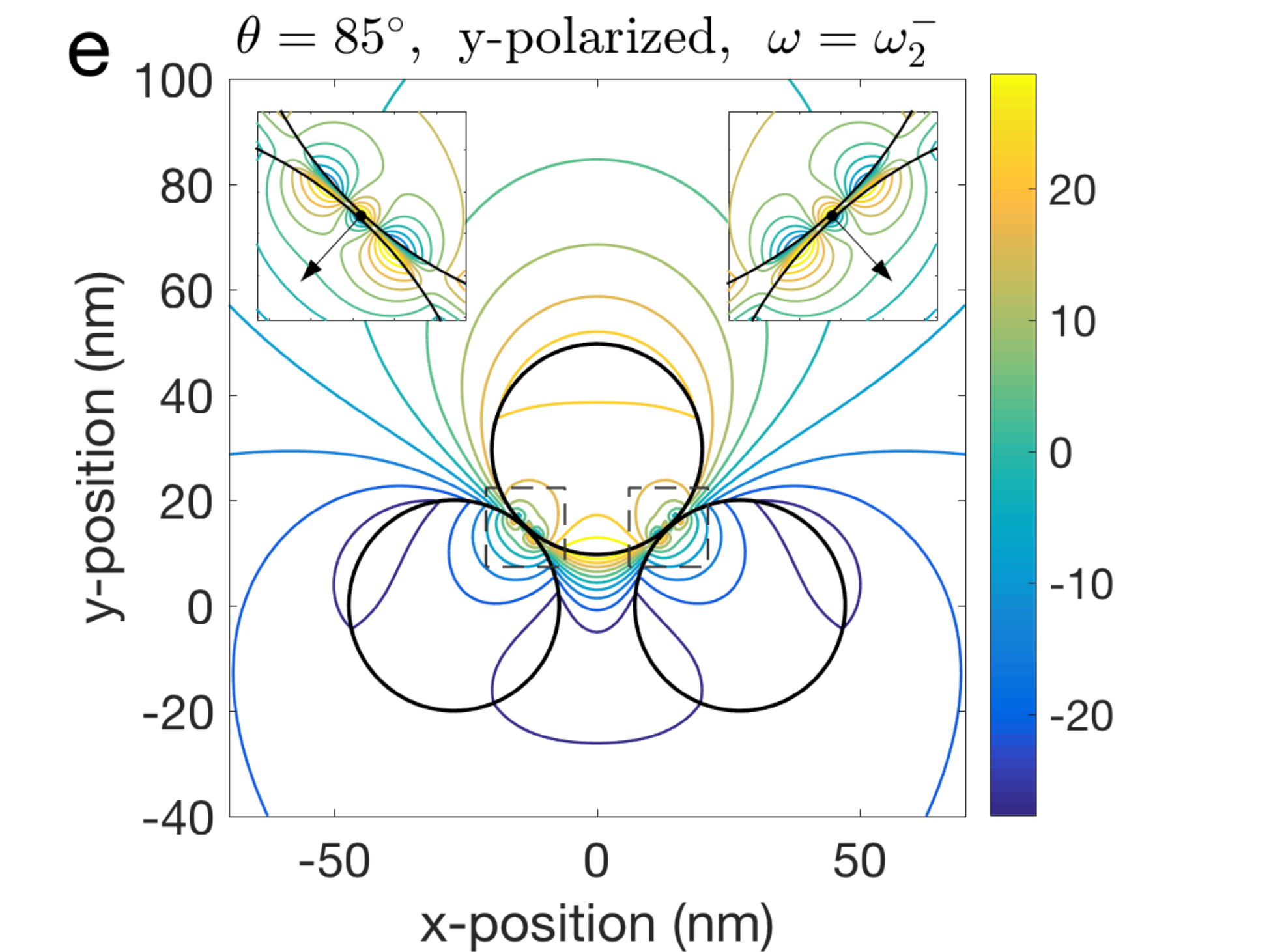}
\\[0.5em]
\includegraphics[width=7.5cm]{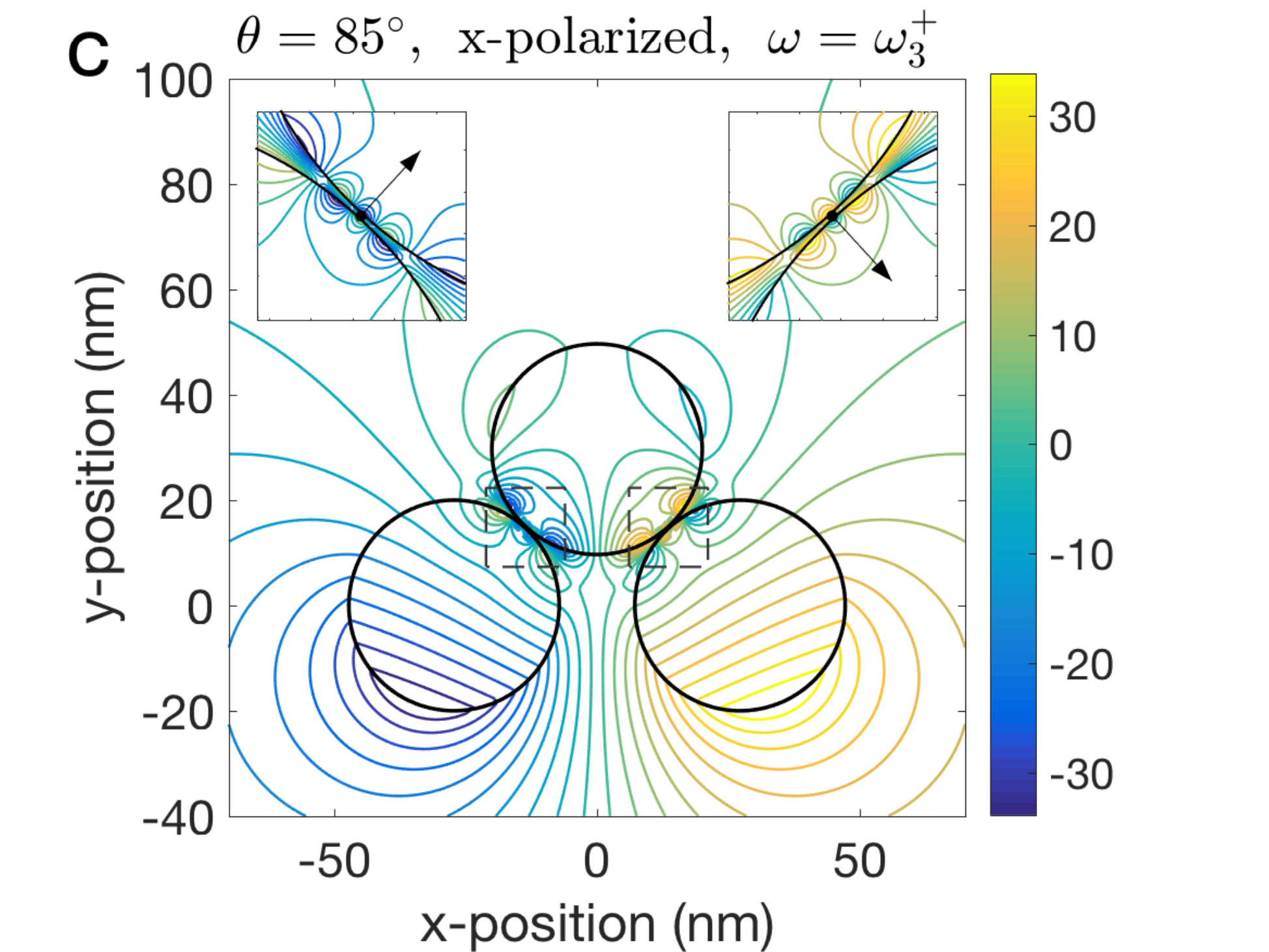}
\hskip-.5cm
\includegraphics[width=7.5cm]{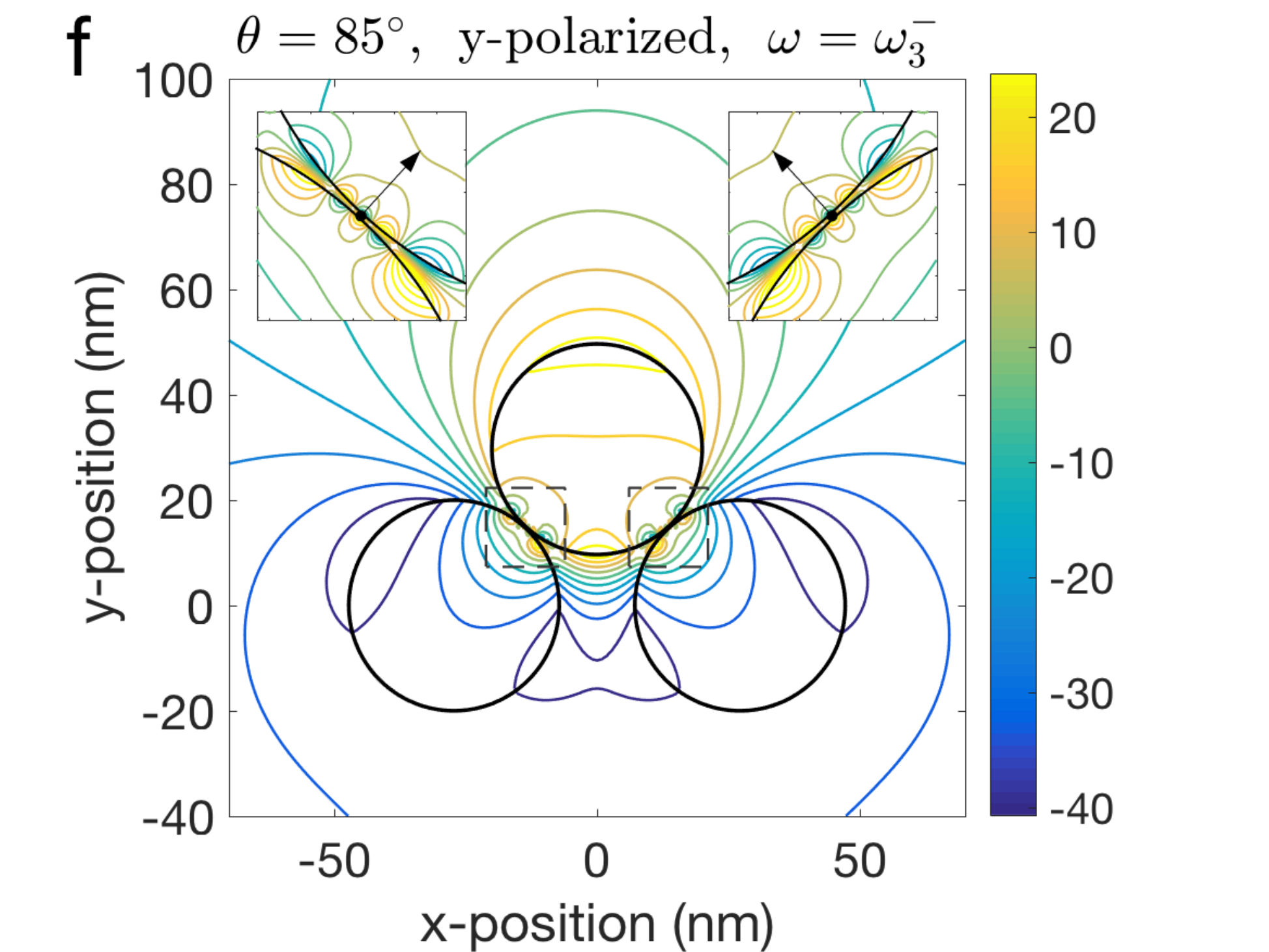}
\caption{
The electric potentials for the trimer with the gap distance $\delta = 0.25$ nm and the bonding angle $85^\circ$ when (A,B,C) the frequency $\omega$ is $\omega = \omega_1^+, \omega_2^+, \omega_3^+$ and the $x$-polarized light is incident; (B) the frequency $\omega$ is set as $\omega = \omega_1^-, \omega_2^-, \omega_3^-$ and the $y$-polarized light is incident. (small insets) Zoomed plots on the gap regions for the pairs $(B_1, B_2)$ and $(B_2,B_3)$. The black arrows mean the electric fields at the gap centers. We set $R = 20$ nm, $\omega_p = 3.85$ eV and $\gamma = 0.1$ eV.
}
\label{figS:contour4}
\end{figure*}

In Fig. \ref{figS:contour3}, we consider the case when the gap distance is smaller as $\delta = 0.25$ nm and the bonding angle is $\theta = 150^\circ$. We plot the potential distributions for six trimer plasmons $|\omega_1^\pm \rangle, |\omega_2^\pm \rangle$ and $|\omega_3^\pm \rangle$. 
In Fig. \ref{figS:contour4}, we consider the same case but with a smaller bonding angle $\theta=85^\circ$.
Again, the potential distributions are dominated by  anti-bonding (and bonding) combinations of two TO dimer plasmons under the $x$-polarized (and $y$-polarized) incident light.

These results also indicate that the resonance peaks of the absorption spectra for the trimer given in the main text are due to the hybridization of singular gap-plasmons.

\section{ The symmetric trimer case}

Let us assume the trimer is symmeric so that all three gap distances are identical to a small positive number $\delta$.
Then, since the particles $B_1$ and $B_3$ are nearly touching,
we also need to include TO dimer plasmons for the pair $(B_1,B_3)$ unlike the previous case.
The hybridized modes can be approximated by $|\omega_n\rangle = a_n |\omega_n^{TO}(B_1,B_2)\rangle + b_n |\omega_n^{TO}(B_2,B_3)\rangle + c_n |\omega_n^{TO}(B_1,B_3)\rangle$. Using the symmetry of the trimer and the anti-symmetry of the TO dimer plasmon $|\omega_n^{TO}\rangle$, one can see from the eigenvalue problem $\mathcal{A}_T|\omega_n\rangle = \omega^2 |\omega_n\rangle$ that the hybridization is (approximately) described by 
\begin{equation}
\label{eq26}
\begin{bmatrix}
(\omega_n^{TO})^2 & \Delta_{n} & - \Delta_n
\\
\Delta_{n} & (\omega_n^{TO})^2 & - \Delta_n
\\
-\Delta_n & - \Delta_n & (\omega_n^{TO})^2
\end{bmatrix}
\begin{bmatrix}
a_n
\\
b_n
\\
c_n
\end{bmatrix}
=
\omega^2
\begin{bmatrix}
a_n
\\
b_n
\\
c_n
\end{bmatrix},
\end{equation}
where $\Delta_n$ is given by Eq. \ref{def_Deltan}.
By solving Eq. \ref{eq26}, we obtain the (approximate) hybridized modes
\begin{align}
    &|\omega_n^1 \rangle \approx \frac{1}{\sqrt{2}}( |\omega_n^{TO}(B_1,B_2)\rangle + |\omega_n^{TO}(B_1,B_3)\rangle ),
    \\
    &|\omega_n^2 \rangle \approx \frac{1}{\sqrt{2}}( |\omega_n^{TO}(B_1,B_2)\rangle - |\omega_n^{TO}(B_2,B_3)\rangle ),
    \\
    &|\omega_n^3 \rangle \approx \frac{1}{\sqrt{3}}( |\omega_n^{TO}(B_1,B_2)\rangle + |\omega_n^{TO}(B_2,B_3)\rangle - |\omega_n^{TO}(B_1,B_3)\rangle ),
\end{align}
and their (approximate) resonant frequencies 
\begin{equation}
    \omega_n^1 \approx \sqrt{(\omega_n^{TO})^2 - \Delta_n}, \quad \omega_n^2 \approx \sqrt{(\omega_n^{TO})^2 - \Delta_n}, \quad \omega_n^3 \approx \sqrt{(\omega_n^{TO})^2 + 2\Delta_n}.
\end{equation}

\section{ SPH theory for the metasurface}

\begin{figure*}
\centering
\includegraphics[width=10cm]{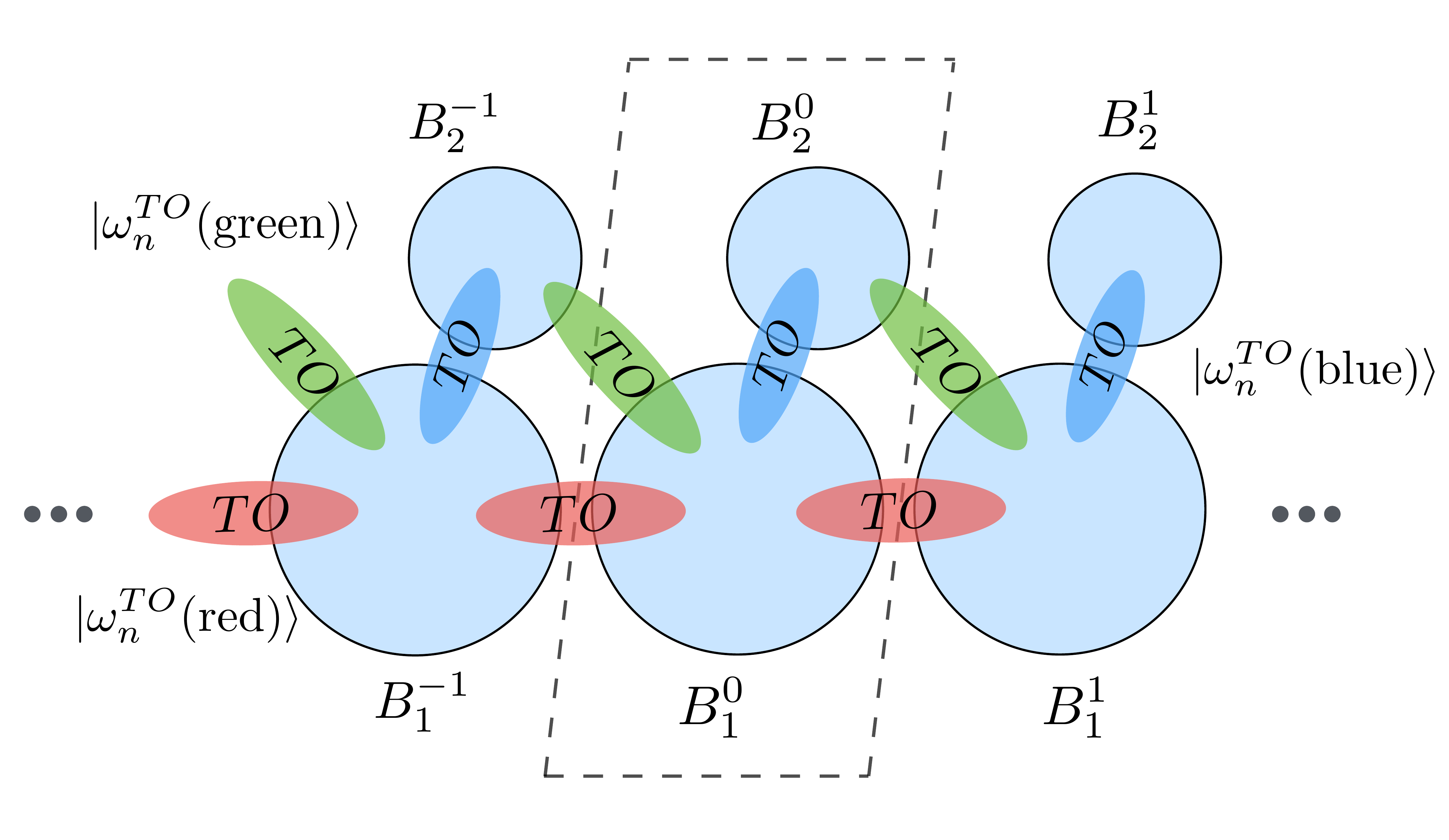}
\caption{
Metasurface and the periodic singular plasmons. The dotted lines represent the unit cell of the metasurface. The red, blue, green coloured arrays of ellipses represent the periodic singular plasmons basis $|\omega_n^{TO}(red)\rangle$, $|\omega_n^{TO}(blue)\rangle$, $|\omega_n^{TO}(green)\rangle$, respectively
}
\label{figS:metasurface}
\end{figure*}

We briefly describe how SPH model can be applied to the metasurface proposed in the main text.
The unit cell of the metasurface consists of two particles. See Fig. \ref{figS:metasurface}. We denote the lower and the upper particles by $B_1^0$ and $B_2^0$, respectively.  The metasurface is periodic in the $x$-direction with the period  $L=2R_1 +\delta_1$. We denote the two particles in the $m$-th translated unit cell by $B_1^m$ and $B_2^m$. More precisely, we set $B_{1}^m = B_{1}^0 +m (L,0)$ and $B_{1}^m = B_{1}^0 +m (L,0)$ for $m\in\mathbb{Z}$.
 
 We introduce the one-dimensional periodic Green's function $G^\#$ for $\mathbb{R}^2$ satisfying
\begin{equation}
-\nabla^2 G^\#(\mathbf{r}) = \sum_{n\in \mathbb{Z}} \delta \big(\mathbf{r} + (n L,0)\big).    
\end{equation}
Here, $\delta$ means the Dirac delta function.
Then it can be shown that
\begin{equation}
    G^\# (\mathbf{r})= -\frac{1}{4\pi L^2} \log ( \sinh^2 \frac{\pi y}{L} + \sin^2 \frac{\pi x}{L} ).
\end{equation}
We also let $\Omega=B_1^0 \cup B_2^0$.
Then, assuming the normal incident field, the charge density $\sigma$ induced on the surfaces $\partial \Omega$ of the particles $B_1^0 \cup B_2^0$ in the unit cell can be determined by
\begin{equation}\label{periodic_integral_eqn}
(\mathcal{K}^{*,\#}_{\Omega} - \lambda I)[\sigma] = \mathbf{E}^{in}\cdot \mathbf{n}|_{\partial\Omega}, \quad \lambda = \frac{\epsilon+1}{2(\epsilon-1)}.
\end{equation}
Here, $\mathcal{K}_{\Omega}^{*,\#}$ is the periodic NP operator defined by
\begin{equation}
    \mathcal{K}_{\Omega}^{*,\#}[\sigma] = -\int_{\partial \Omega } \frac{\partial G^\#(\mathbf{r}-\mathbf{r}')}{\partial \mathbf{n}(\mathbf{r})} \sigma (\mathbf{r}') d S(\mathbf{r}'), \quad \mathbf{r} \in \partial \Omega.
\end{equation}
We should mention that the above equation (\ref{periodic_integral_eqn}) is derived based on the quasi-static approximation, hence it is accurate only when the period $L$ is much smaller than the wavelength.

We define $\mathcal{A}_{\Omega}^\#$ by
\begin{equation}
    \mathcal{A}_{\Omega }^{\#}[\sigma] = \omega_p^2 \left( \frac{1}{2}I - \mathcal{K}_{\Omega }^{*,\#} \right)[\sigma].
\end{equation}
Then, as before, the (periodic) surface plasmons of the structure can be determined by the eigenvalues and eigenfunctions of the operator $\mathcal{A}_{\Omega}^\#$.

We define three kinds of periodic arrays of gap plasmons by
\begin{align}
   & |\omega_n^{TO}(\mbox{red})\rangle :=\sum_{m\in\mathbb{Z}}|\omega_n^{TO}(B_1^m, B_1^{m+1})\rangle, \\
   & |\omega_n^{TO}(\mbox{blue})\rangle :=\sum_{m\in\mathbb{Z}}|\omega_n^{TO}(B_1^m, B_2^{m})\rangle, \\
   & |\omega_n^{TO}(\mbox{green})\rangle :=\sum_{m\in\mathbb{Z}}|\omega_n^{TO}(B_1^m, B_2^{m+1})\rangle.
\end{align}
Using these singular gap plasmons as basis, we can construct a matrix representation of the operator $\mathcal{A}^\#_\Omega$ analytically as in the trimer case.
When doing so, one should restrict each periodic singular plasmon onto the surfaces $\partial \Omega=\partial B_1^0 \cup \partial B_2^0$ of the particles in the unit cell.
By restricting these plasmons on the surfaces $\partial\Omega$ of the particles in the unit cell, we construct good basis for the hybridized modes of the metasurface. The more detailed analysis on the hybridized modes is  non-trivial and will be considered elsewhere.
Once we compute the eigenvalues $\{(\omega_j^\#)^2\}_{j=1}^\infty$ and eigenfunctions $\{| \omega_j^\#  \rangle\}_{j=1}^\infty$ of the operator $\mathcal{A}_{\Omega}^\#$, as in the trimer case, we can determine the charge density $\sigma$ by
\begin{equation}
    \sigma = \sum_{j=1}^\infty \frac{\omega_p^2\langle  \mathbf{E}^{in}\cdot \mathbf{n}|_{\partial \Omega}|\omega_j^\#\rangle}{\omega^2 - \omega_n^2 + i \gamma \omega} |\omega_j^\#\rangle, 
\end{equation}
from which we obtain its dipole moment $\mathbf{p}_\Omega = (p_{\Omega,x},p_{\Omega,y})$. Then the reflection $\mathcal{R}$, transmission $\mathcal{T}$ and absorption $\mathcal{A}$ can be computed (approximately) as follows \cite{PendryPRX2015}:
\begin{equation}
    \mathcal{R} = |r_{meta}|^2, \quad \mathcal{T} = |t_{meta}|^2, \quad \mathcal{A}= 1- \mathcal{R}-\mathcal{T},
\end{equation}
where
\begin{equation}
   r_{meta} = - \frac{k_0 J/\omega}{2 + k_0 J/ \omega}, \quad
   t_{meta} = \frac{2}{2+k_0 J/\omega}, \quad  J = - \frac{i \omega p_{\Omega,x}}{L}.
\end{equation}

%\newpage

%\newpage

\bibliographystyle{IEEEtran}
\bibliography{arXiv_FINAL}
\end{document}